\newcommand{\Ks}{$\mathrm{K_S}$}
\newcommand{\J}{$\mathrm{J}$}

\documentclass{aa}  

\usepackage{graphicx}
\usepackage{txfonts}
\usepackage{hyperref}
\usepackage{xcolor}
\usepackage[normalem]{ulem}
\usepackage{booktabs}
\begin{document} 

    \title{VVV CL001: A strong candidate for the most ancient star cluster in the Galaxy}


\author{  J.~Olivares~Carvajal\inst{1,2}
          \and
          Á.~Valenzuela~Navarro\inst{2}
          \and
          C.~Salvo-Guajardo\inst{2}
          \and
          C.~Ordenes-Huanca\inst{3,2}
          \and
          B.~Acosta-Tripailao\inst{2}
          \and
          C.~Quezada\inst{2}
          \and
          F.~Gran\inst{4,2}
          \and
          M.~Zoccali\inst{2}
          \and
          M.~Dall'Ora\inst{1}
          \and
          M.~De~Leo\inst{5,6}
          \and
          S.~Villanova\inst{7}
          \and 
          D.~Geisler\inst{8,3}
          }

 \institute{INAF - Osservatorio Astronomico di Capodimonte, Salita Moiariello 16, 80131, Naples, Italy\\
              \email{julio.olivarescarvajal@inaf.it}
         \and
         Instituto de Astrof\'isica, Pontificia Universidad Cat\'olica de Chile, Av. Vicu\~na Mackenna 4860, 782-0436 Macul, Santiago, Chile
         \and
             Departamento de Astronomía, Universidad de Concepción, Casilla 160-C, Concepción, Chile
         \and 
         Gemini Observatory/NSF’s NOIRLab, Casilla 603, La Serena, Chile
         \and 
         Dipartimento di Fisica e Astronomia, Università degli Studi di Bologna, Via Piero Gobetti 93/2, Bologna, 40129, Italy
         \and
            Osservatorio di Astrofisica e Scienza dello Spazio di Bologna, INAF, Via Piero Gobetti 93/3, Bologna, 40129, Italy         
            \and
            Universidad Andrés Bello, Facultad de Ciencias Exactas, Departamento de Física y Astronomía - Instituto de Astrofísica,
Autopista Concepción - Talcahuano 7100, Talcahuano, Chile
            \and
            Departamento de Astronomía, Facultad de Ciencias, Universidad de La Serena. Av. Raúl Bitrán 1305 , La Serena, Chile
            }
   \date{Received 2026, September 16; accepted 2026, October 20}

 
  \abstract
   {Finding the oldest globular cluster of the Galaxy is important to constrain a lower limit on the age of the Galaxy and the Universe. VVV CL001 $([l,b]=[5.26^\circ,0.77^\circ])$ is perhaps the most metal-poor ($\rm [Fe/H]\sim-2.2 \ dex$) globular cluster in the Galactic bulge and a promising candidate to be the oldest globular cluster in the entire Milky Way. Due to its recent discovery and being behind a thick veil of interstellar clouds, its physical parameters, especially the distance and age, remain highly uncertain, yet they are essential to define it as an ancient native stellar system of the Milky Way.}
   {We aim to refine the distance and age of VVV CL001 using deep and high-quality photometry, to study its role in the formation of the Milky Way, and to determine whether it is the oldest Galactic globular cluster.}
   {We used \J\ and \Ks\ observations to construct the deepest CMD of
VVV CL001 to date, based on FourStar/Baade PSF photometry combined with HST
near-IR wide-filter PSF photometry. We selected cluster members through proper
motion calculations and the HDBSCAN algorithm, and assigned membership
probabilities via Markov-Chain Monte Carlo iterations in a 5-dimensional
parameter space. Using $r$ band data from DECam/Blanco, we optically identified
horizontal branch members and applied an isochrone-fitting technique to
determine the cluster distance, while members around the main-sequence turn-off
yielded the age from different isochrone sets. Finally, we refined the cluster
orbital parameters.}
   {We obtained a distance of $d=7.34\pm0.46$ kpc, the most precise distance
reported in the literature so far. We derived an age of $13.4\pm0.4$ Gyr,
indicating that VVV CL001 is one of the oldest globular clusters in the Galaxy.}
   {The orbit confirms that the cluster belongs to the Galactic bulge, which,
together with its age and metallicity, could link it to the early proto-Milky Way assembly, extending the metal-poor end of the known {\it in situ} globular clusters.}
   \keywords{Galaxy: bulge -- globular cluster: individual: VVV CL001 --  Galaxy: formation -- Galaxy: disk}

   \maketitle
%

\section{Introduction} \label{sec:intro}

Globular clusters (GCs) are among the oldest fossils of the Milky Way (MW), preserving a direct record of the conditions under which our Galaxy first assembled~\citep{Kruijssen2019}. These clusters consist of old stars with ages ranging from $10$ to more than $13$~Gyr~\citep{ying+2023, valcin+2025} that were born from the same gas cloud and, therefore, share similar chemical compositions~\citep{Gratton2019}. As a result, these stars can be considered a simple stellar population, helping to estimate ages and distances within the Galaxy. They are regularly used to calibrate chemo-dynamical tracers or chemo-dynamical relations, thus playing a crucial role in Galactic distance scales~\citep[for instance,][]{Bhardwaj2023}.

The MW hosts approximately $170$ GCs to date~\citep{Harris2010,Baumgardt2018}. A large number of studies related to GCs have been providing important constraints in the different properties of the Galaxy: ages, distances, dynamics, chemical abundances and merger history \citep{Kruijssen2020}, helping in the complete characterization of the MW. Several models predict that, close to the center of the Galaxy, we should find its oldest GC, the so-called oldest native GC of the MW, or the the Main Progenitor~\citep{Geisler2025}, which can be even older than those in the halo~\citep{Barbuy2006,Barbuy2018,Savino2020}. That object, and its age, would allow us to assign a lower age limit to the Galaxy, and to the Universe. This will provide important insights to the current debate about the Hubble constant $H_0$~\citep{CimattiMoresco2023}. 

Because the cluster metallicity is much more straightforward to measure than the absolute age, several authors have selected the most metal-poor bulge GCs as candidate oldest native clusters and have tried to derive precise ages for them. The task is non trivial, as this region is characterized by high extinction, stellar crowding, and significant foreground contamination. A well-studied relatively metal-poor bulge GC is NGC 6558, for which~\cite{Barbuy2018} and~\cite{Souza2024} measured an age of $13.2$ Gyr, confirming the presence of very old GCs in the bulge. Very recently, \cite{Ortolani2026} studied the bulge relatively metal-rich GC Ton 2, and estimated an age of $13.6$ Gyr. Everything else being equal, one expects that the oldest GC should be among the most metal-poor ones \citep{Geisler2025}.

\begin{figure*}[ht!]
\centering
\includegraphics[width=\textwidth]{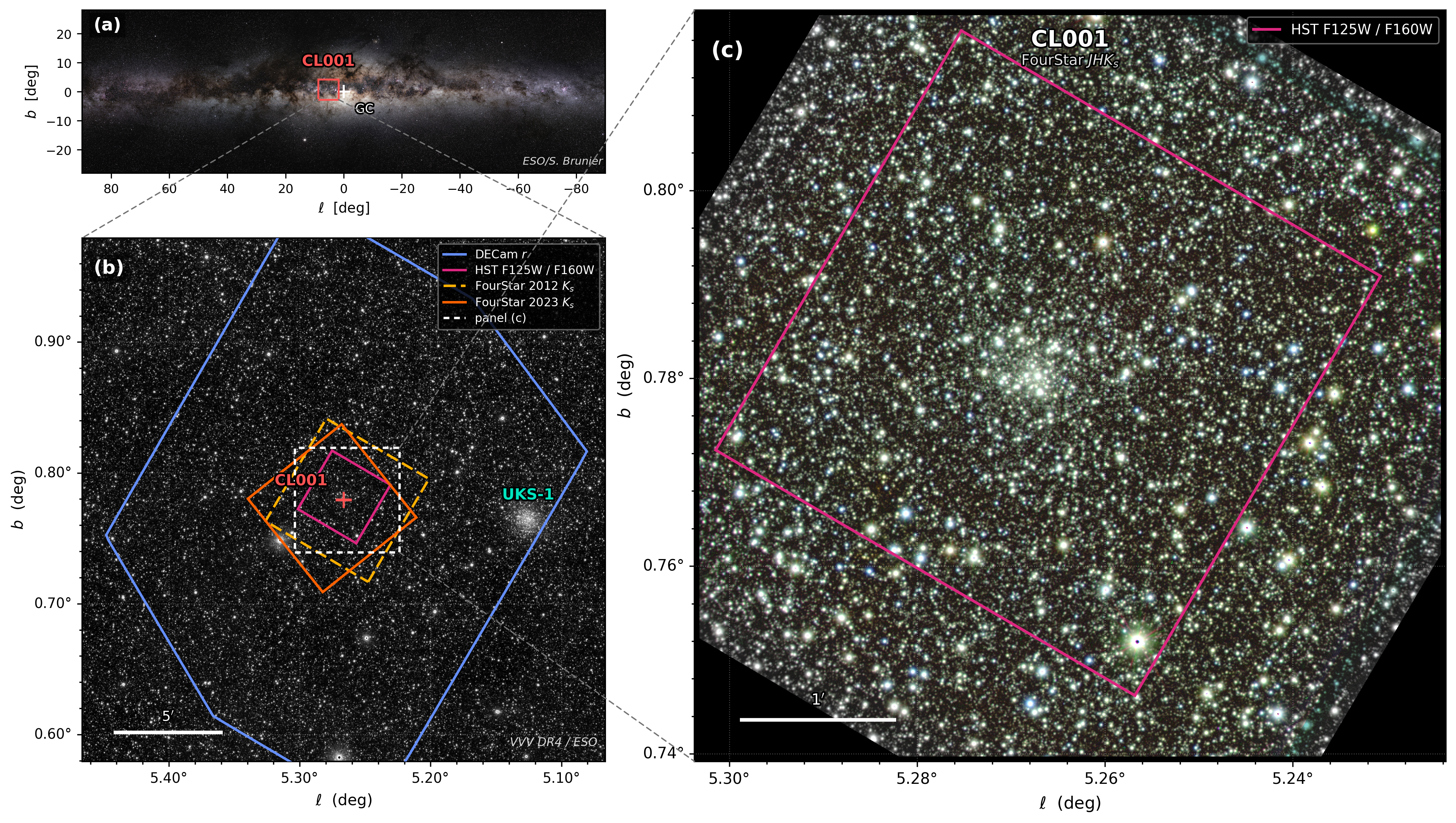}
  \caption{Location of CL001. \textit{(a)} Optical panorama of the Milky Way (ESO/S. Brunier) in Galactic coordinates. The box marks CL001 at $(\ell, b) = (5.27^\circ, +0.78^\circ)$ and is enlarged for visibility. \textit{(b)} $24'$ VVV DR4 H-band field centred on CL001, with the neighbouring massive cluster UKS-1 $8.5'$ away, showing the footprints of the data used in this work: DECam $r$ (blue, extending beyond the panel), FourStar \Ks \ 2012 (gold dashed) and 2023 (red solid), the two epochs are rotated one with respect to the other), and HST/WFC3-IR (pink). The dashed white box is the field of view of panel (c). \textit{(c)} FourStar $JHK_s$ color composite of CL001 with the HST footprint. All panels are in Galactic coordinates.}
  \label{fig:RGBimage}
\end{figure*}

An exceptionally metal-poor GC example is the so-called VVV CL001~\citep{Minniti2011}, which was first discovered by the VISTA Variables in the Vía Láctea Survey~\citep[VVV,][]{Minitti2010}. The cluster is embedded in a highly extincted region of the bulge ($A_V\sim9.3$ mag); thus, only NIR observations have been used to study it. It was observed as a low-mass, low-luminosity stellar group in the inner bulge region, as observed in Fig.~\ref{fig:RGBimage} (a), and was dynamically confirmed as a cluster using its proper motions (PMs) in \cite{Gran2019}. Later, \citet[][hereafter FT21]{FT2021} found two stars associated with the cluster using APOGEE-2S spectra~\citep{APOGEEDR2}, and \citet[][hereafter OC22]{OC2022} derived radial velocities (RVs) and metallicities by MUSE spectra~\citep{MUSE} observations of all the stars within 1x1 arcmin from the center, and confirmed 56 cluster members. Recently, \citet[][hereafter HM26]{moya_2026} derived chemical abundances of 12 cluster members by means of a FORS2 calcium triplet spectroscopic analysis. These previous studies also argued for a possible ancient origin of the cluster, given its very low metallicity, with [Fe/H] $\approx-2.2$ dex. Nonetheless, the cluster's exact age remains poorly constrained.

Despite the fact that the VVV CL001 projected sky position is very close to the MW center, its mean RV ($-325$ km/s) is markedly different from that of bulge field stars suggesting that the cluster could not be part of the bulge. Therefore, determining the cluster's orbit is essential to understanding its origin. FT21 attempted to obtain an orbit assuming a distance of 8.5 kpc, though with a rather large error of $\pm2$ kpc ($\sim 23\%$). If we set the distance to the upper limit, the resulting orbit would not lie within the bulge. Conversely, OC22 used a distance of $8.23$ kpc and an upper limit of $9.1$ kpc; in most cases, the cluster orbit lay within the bulge. HM26 obtained a distance of $7.1$ kpc with an error of $1$ kpc; in all cases, the cluster's orbit is within the old disk and the bulge. However, \cite{Geisler2025}, although classifying it officially as a bulge GC based on a reassessment of a variety of chemical and dynamical studies, suggested it was most likely a disc GC due to its extreme metallicity, being the most metal-poor GC in their list of bulge GCs. In summary, the differences in orbital parameters reported in previous studies are mainly driven by uncertainty in the distance.

Unfortunately, none of the previous studies have measured the cluster's distance with an error of less than $\pm1$ kpc, as it is located in a very crowded, highly extincted region of the bulge and is a very faint, low-mass object. In fact, the color-magnitude diagram (CMD) from previous studies is sparsely populated and rather shallow. It had no confirmed members in the horizontal branch (HB) and did not reach the main sequence turnoff (TO). Moreover, no RR Lyrae stars have been confirmed as members of the cluster so far, making it impossible to calculate the distance by this method~\citep{OC2022,Zoccali2024}. Therefore, the distance error is the primary factor contributing to orbit uncertainties. Deeper photometry is needed to identify additional cluster members and reduce this distance error, thereby enabling estimation of the cluster age.

In this work, we present deep near-infrared (NIR) observations of the VVV CL001 cluster with FourStar on the Magellan Baade 6.5m Telescope, combined with optical $r$ band photometry from archival data of the DECam/Blanco 4m telescope, to precisely determine the cluster distance. We complement our observations with previous FourStar images to obtain precise PMs for the cluster selection. We use new Hubble Space Telescope (HST) observations to determine the cluster age. The paper structure is as follows: Section~\ref{sec:data} describes the data used, the different photometry, and calibrations. Section~\ref{sec:PMs} presents the method employed to derive the PMs. Section~\ref{sec:members} describes the selection of cluster members. In Section~\ref{sec:distage}, we present the results for the cluster's distance and age. Section~\ref{sec:orbit} shows the orbital integration. Finally, in Section~\ref{sec:discussion} and \ref{sec:conclusions}, we present our discussion and conclusions, respectively.   

\section{Data} \label{sec:data}

\subsection{Baade Telescope - FourStar}

We used J, H, and \Ks \ observations using the NIR imager FourStar~\citep{Persson+2013}, located at the 6.5 m Baade Telescope, at Las Campanas Observatory. FourStar is composed of four detectors with a total field of view of $10.9' \times 10.9'$ and a pixel scale of $0.159''$. The observations were conducted on the nights of May 5, 2012~\citep[hereafter, Epoch 1,][]{Haro-Moya+2025}, and on May 7, 2023 (hereafter, Epoch 2). The reduction of the FourStar images was carried out using the \texttt{THELI} pipeline~\citep{erben_2005, Schirmer2013}, comprising dark subtraction, flat-field correction, and sky background modeling and subtraction.

We performed PSF photometry using the \texttt{DAOPHOT/ALLSTAR} and \texttt{ALLFRAME} suite~\citep{stetson1987,stetson1994}. We stacked all images from Epochs 1 and 2 in all filters to create a master list of stars. Afterward, we selected stars to model the PSF in the stacked image and then computed a PSF model from these stars for each individual image. We applied \texttt{ALLFRAME} to each epoch and obtained a PSF-photometric catalog for each epoch. The astrometric calibration is discussed in Sec.~\ref{sec:astrometry}. Panel c of Fig.~\ref{fig:RGBimage} shows the color $JHK_s$ composite image of VVV CL001 using FourStar data.

\subsection{Hubble Space Telescope - WFC3/IR} \label{subsec:hst}

We used archival images from HST in filters F125W and F160W with WFC3/IR (\citet{massari+2023}, proposal ID 17435, hereafter referred to as Epoch 3), observed on February 18, 2025\footnote{Photometric catalogs are publicly available. See \citet{libralato+2026}.}. This provides a temporal baseline of $12.79$ years between the Epochs 1 and 3, and $11.01$ years between the Epochs 1 and 2. This large temporal baseline is ideal for accurately computing PMs.\\

For the HST photometry in the F125W and F160W filters, we used \texttt{DOLPHOT}, which was specifically developed and optimized for high-precision photometry space-based observations~\citep{Dolphin2000, Dolphin2016}. Briefly, we ran \texttt{DOLPHOT} using the parameters recommended by \citet{Dolphin2000, Dolphin2016} for highly crowded fields. We used the drizzled F160W image as the reference frame for source detection and to define the common astrometric system, while the final magnitudes in both filters were obtained via simultaneous PSF fitting of the individually calibrated exposures (FLT images). This strategy allows the drizzled image to be used for registration and source detection while avoiding the resampling effects introduced by the drizzle process in the photometry, since the PSF fitting is performed directly on the individual exposures.

\subsection{Blanco 4~m Telescope - DECam}

For the $r$ band DECam observations, we performed PSF photometry using \texttt{DAOPHOT}. For each individual epoch, we ran \texttt{DAOFIND} and \texttt{ALLSTAR} to detect sources and obtain PSF photometry. The resulting source catalogs were then processed through \texttt{DAOMATCH} and \texttt{DAOMASTER} to establish a common astrometric reference frame and assign consistent stellar IDs across all epochs. Finally, we cross-matched the individual-epoch catalogs and computed the mean $r$ band magnitude for each source. Further details on the DECam photometry are provided in \cite{Gholami2026}. On Fig.~\ref{fig:RGBimage} (b), there is a comparison of the field-of-views of the different datasets used in this work. 

\begin{figure*}
\centering
\includegraphics[width=16cm]{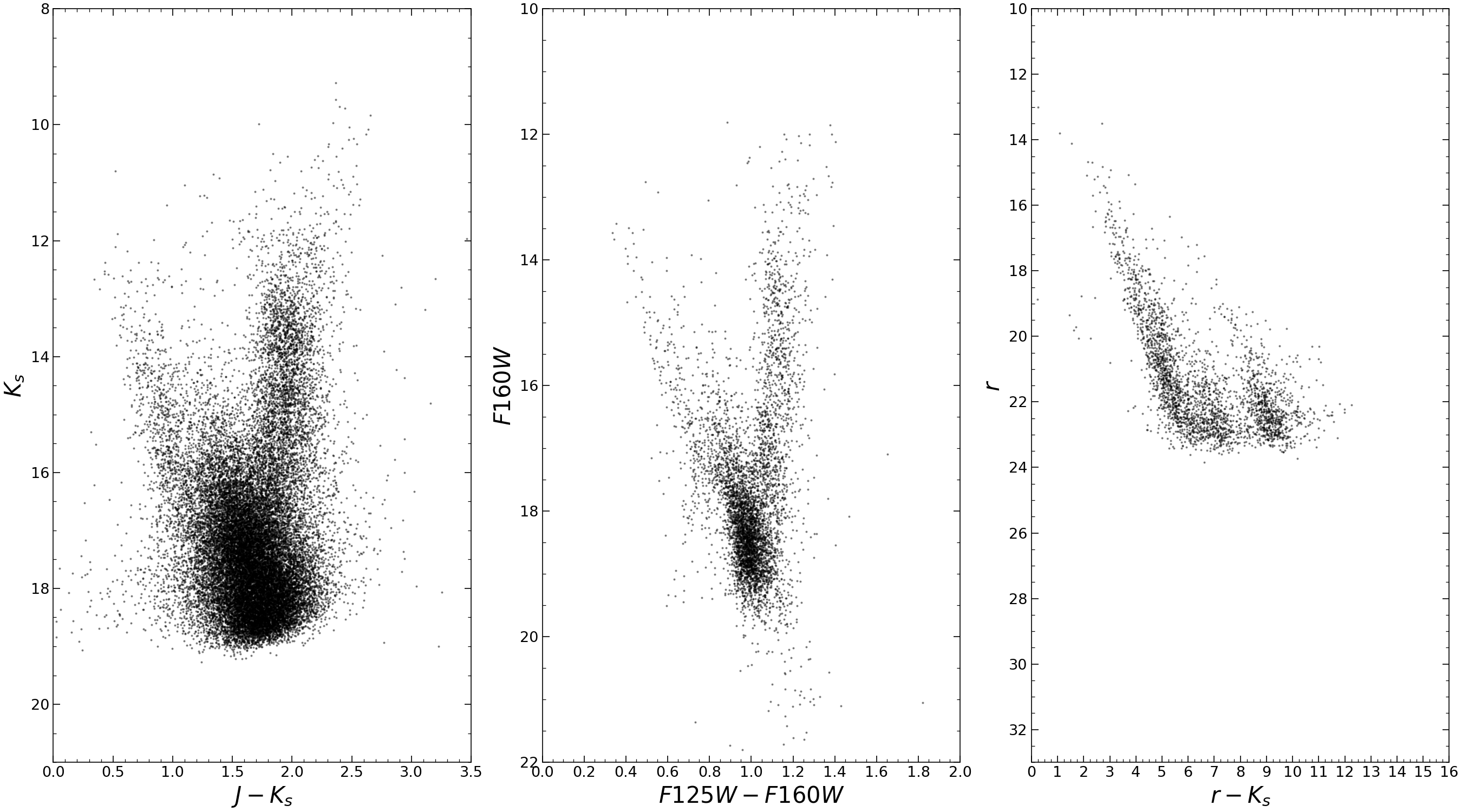}
  \caption{The different photometry available for the VVV CL001 field. \textit{Left panel:} The NIR FourStar photometry based on the \J \ and \Ks \ bands. \textit{Middle panel:} The NIR HST wide field photometry based on the $F125W$ and $F160W$ bands. \textit{Right panel:} The CMD showing the optical DECam $r$ band photometry and the NIR FourStar \Ks \ band. It shows fewer stars due to the high extinction in the field. }
     \label{fig:cmds}
\end{figure*}

\subsection{Photometric calibration}

Aiming to calibrate the FourStar photometry to the VIRAC2~\citep{smith2025} system, we made a cross-match in \texttt{Topcat}~\citet{taylor2005,taylor2017} for stars brighter than $\mathrm{K_S} = 13$ to avoid mismatches due to the different depths and spatial resolution of both catalogs. We allowed a maximum offset of 0.5 arcsec. The magnitude limit was chosen to isolate RGB stars, thereby discarding cluster and bulge main-sequence (MS) stars, yielding 5\,498 stars in common. Then, we checked if there were any color terms (CTs) in the photometry, due to intrinsic differences between the filters. To compute the zero points (ZPs), we plotted the differences between each filter as a function of the colors $\mathrm{J-K_S}$ and $\mathrm{H-K_S}$. Then we applied a 1-sigma clipping to the data to eliminate mismatching stars. Finally, we binned the data into equally spaced color bins, computed the median for each bin, and fitted a first-degree polynomial to the bin medians using least-squares minimization. The slope of this line represents the CT, and the intercept is the ZP. We did not find significant CT in any of the filters, so we applied only the ZP to calibrate the data using the standard linear relation $m_{cal}=m_{inst}+ZP$. The differences between the ZP using $\mathrm{J - K_S}$ and $\mathrm{H - K_S}$ are negligible. The CMD is shown in Fig.~\ref{fig:cmds} left panel.

For the DECam data, photometric ZPs were calculated using Gaia DR3 synthetic photometry~\citep{Montegriffo2023}, obtained by transforming Gaia BP/RP mean spectra into the standardized SDSS System~\citep{gaiadr3}. The observed instrumental magnitudes were transformed to the calibrated system using the same relation presented before for the FourStar data. In this case, we worked only with the $r$ band, so no CT is available, and in previous works, it was negligible. For more details about this photometric calibration, see Luna et al. (\textit{in prep.}).

For the HST data, \texttt{DOLPHOT} directly calibrates the photometry onto the VEGAmag system using the WFC3/IR ZPs provided by STScI in the image headers. Therefore, no additional external photometric calibration was necessary.

\section{Proper motions} \label{sec:PMs}

We measured PMs using only the Ks-band exposures from Epochs 1 and 2, and the F160W exposures from Epoch 3, as these provide the deepest observations and maximize the number of sources detected in common across the three epochs. The methodology consisted of four main steps. First, we refined the initial stellar positions within each epoch and corrected the astrometry for geometric distortions. Second, we transformed the positions in each exposure onto the Gaia\,DR3 reference frame. Third, we estimated the astrometric uncertainty associated with each stellar position. Finally, we derived PMs by performing linear fits to the positions as a function of time.

\subsection{Astrometric refinement and distortion correction}\label{sec:astrometry}

For each epoch independently, we first ran \texttt{ALLFRAME}~\citep{stetson1994} to refine the stellar centroids and ensure a homogeneous astrometric treatment across all three epochs. For Epoch 3, although the final HST photometry adopted throughout this work was obtained with \texttt{DOLPHOT} (Sect.~\ref{subsec:hst}), the F160W exposures were independently processed with \texttt{DAOPHOT/ALLFRAME} solely for the PM analysis. This allowed the stellar positions to be measured consistently with those of the FourStar epochs, while the \texttt{DOLPHOT} magnitudes were retained for all photometric analyses. We then used \texttt{SCAMP}~\citep{bertin2006} on each epoch separately to compute astrometric solutions corrected for geometric distortions in each individual observation. The refined \texttt{SCAMP} astrometric solutions were used to transform the pixel coordinates (X, Y) from the \texttt{ALLFRAME} catalogs into the initial equatorial coordinates ($\alpha,\delta$) used for cross-matching with Gaia\,DR3.\\

\subsection{Calibration to the Gaia reference frame}

We constructed a reference catalog from Gaia\,DR3 sources by applying quality cuts to ensure astrometric reliability. Following \citet{lindegren+2021}, we only retained sources with $\mathrm{astrometric\_params\_solved=31}$ (5-parameter solution), $\mathrm{duplicated\_source}=\mathrm{False}$, $\mathrm{astrometric\_excess\_noise\_sig}\leq2$, $\mathrm{G}>13$, and $\mathrm{RUWE}<1.4$. As Gaia\,DR3 reference epoch is 2016.0, for each observation, we propagated the Gaia positions to the observation epoch of each observation using Gaia PMs.

To calibrate the astrometry of each frame onto the Gaia\,DR3 reference system, we cross-matched our distortion-corrected catalogs with the epoch-propagated Gaia catalogs using a matching radius of 0.3~arcsec, retaining only sources with photometric uncertainties smaller than 0.10~mag. Following \citet{smith2025} and Salvo-Guajardo et al. (\textit{submitted to A\&A}), we modeled the transformation between the measured pixel coordinates $(X,Y)$ and the Gaia tangent-plane coordinates $(\xi,\eta)$ using fourth-order Chebyshev polynomials. The polynomial coefficients were determined independently for each frame using the matched Gaia sources. We tested higher-order transformations, but these did not produce a significant improvement in the astrometric residuals and were therefore not adopted to avoid potential overfitting. The resulting calibrated tangent-plane coordinates were subsequently projected onto equatorial coordinates $(\alpha,\delta)$ in the Gaia\,DR3 reference frame. Further details are discussed in Salvo-Guajardo et al. (\textit{submitted to A\&A}).

The final calibrated positions reduced the RMS of the astrometric residuals with respect to Gaia\,DR3 from $\sim$0.2-0.3 arcsec before calibration to $\sim$0.01-0.02 arcsec after calibration, corresponding to an improvement of more than 90\% for all frames.

\begin{figure*}[h!]
\centering
\includegraphics[width=1\hsize]{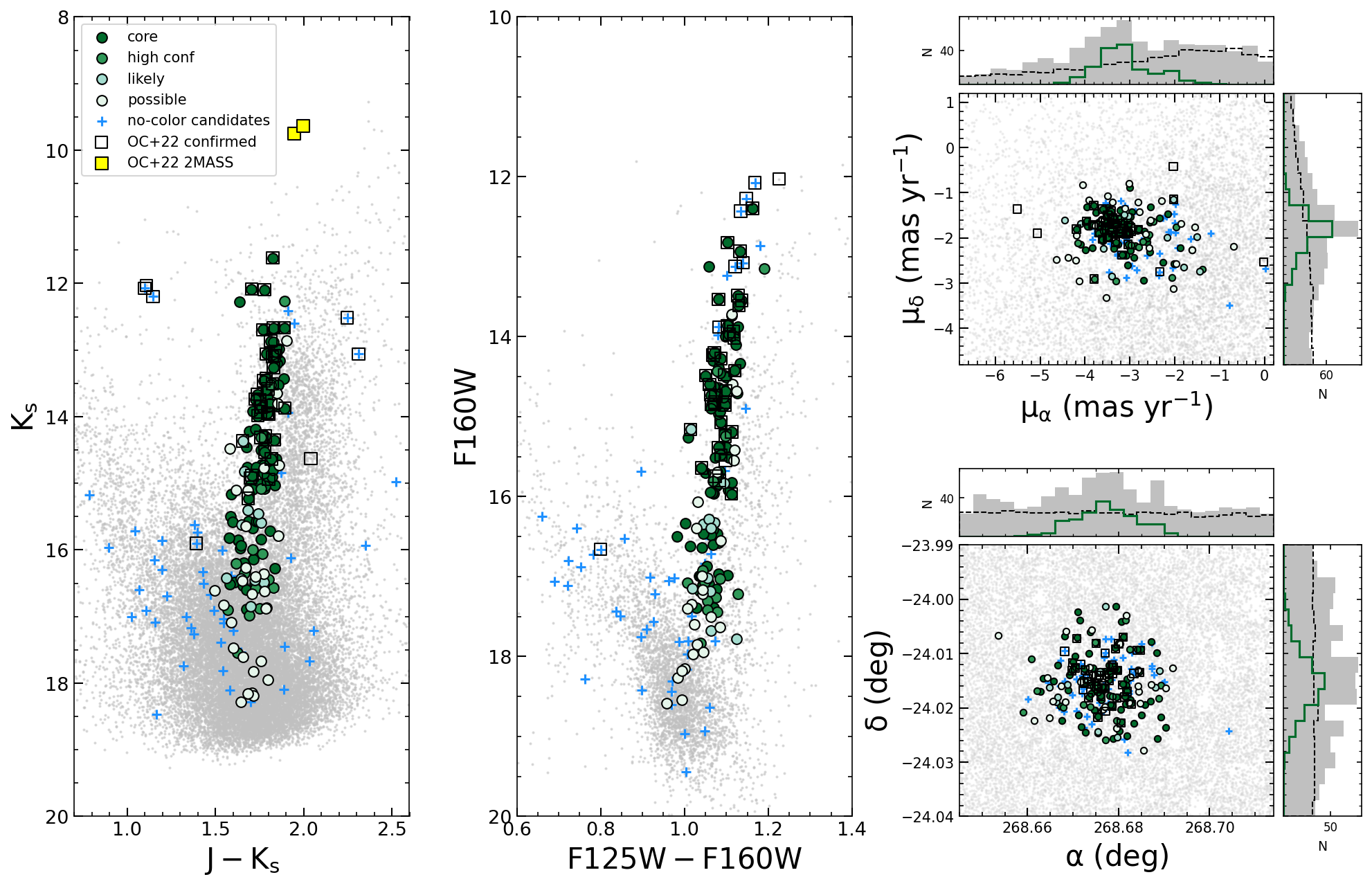}
  \caption{Classification of the members of VVV CL001. \textit{Left panel:} FourStar $K_{\rm s}$ vs. $J-K_{\rm s}$ CMD with the members color-coded by membership class, from dark to light green: \textit{core}, \textit{high confidence}, \textit{likely}, and \textit{possible} members. Black squares are spectroscopically confirmed members from OC22 using MUSE, with VVV photometry. The two yellow boxes are also from OC22 but with photometry from 2MASS. Members selected through the four-dimensional parameter space procedure are marked as blue crosses. Gray points are field stars. \textit{Middle panel:} Same as left panel, but for the HST bands. \textit{Upper right panel:} VPD of our selected members compared to field stars. The marginal histograms show the PM distribution of all stars within $45''$ of the cluster center (gray), the field contribution expected in that area, estimated from stars beyond $100''$ (dashed line), and the cluster members in the same area (green). \textit{Lower right panel:} Spatial distribution of all stars in the same field of CL001 and the selected cluster members.}
     \label{fig:members}
\end{figure*}

\subsection{Positional uncertainty estimation}

Since \texttt{DAOPHOT} and \texttt{ALLFRAME} do not provide individual astrometric uncertainties for each detection, we derived approximate positional uncertainties from photometric errors. Following the approach of Salvo-Guajardo et al. (\textit{in prep.}), based on the formalism of Lindegren (1978), the astrometric uncertainty can be directly approximated from the photometric uncertainty as $\sigma_{\mathrm{ast}} \approx 0.39 \times \mathrm{FWHM} \times \sigma_{\mathrm{phot}}$. Thus, the total positional uncertainty for each measurement ($\sigma_{\mathrm{pos}}$) was computed as the quadrature sum of the individual astrometric uncertainty ($\sigma_{\mathrm{ast}}$) and the frame-level Gaia calibration uncertainty. The latter was taken as the RMS of the positional residuals of the stars used to derive the transformation onto the Gaia\,DR3 reference frame.

\subsection{Proper motion determination}

PMs were derived for all stars detected in Epoch 1 and in at least one additional epoch (either Epoch 2 or Epoch 3) by fitting their calibrated $\alpha^*$ and $\delta$ positions as a function of time. For each coordinate, we adopted a linear model,
\begin{align}
\alpha^*(t) &= \alpha^*_0 + \mu_{\alpha^*}(t - t_0),\\
\delta(t) &= \delta_0 + \mu_{\delta}(t - t_0).
\end{align}
where $(\alpha^*_0, \delta_0)$ are the fitted positions at the reference epoch $t_0$, and $(\mu_{\alpha^*}, \mu_{\delta})$ are the corresponding PM components. The fits were performed using weighted least squares, with each positional measurement weighted by $1/\sigma_{\mathrm{pos}}^2$. Here, $t_0$ was taken as the weighted mean epoch of the observations using the same weights. Finally, the uncertainties in the resulting PMs were obtained directly from the covariance matrix of the fits, leading to median total PM uncertainties of $\sim0.5$~mas\,yr$^{-1}$ for stars with \Ks \ < 16 and $\sim1$~mas\,yr$^{-1}$ for stars with \Ks \ < 18.\\

\begin{table}
\centering
\caption{Summary of the number of stars comprising each defined membership class for CL001.}
\label{tab:membership}
\begin{tabular}{lrrrrl}
\hline \hline
Class & N \\
\midrule
Core            &  $92$\\
High-confidence &  $19^{a}$\\
Likely          &  $11$\\
Possible        &  $33^{b}$\\
\midrule
Total           & $155$\\
\hline \hline
\end{tabular}
\tablefoot{$^{a}$ includes the two confirmed HB stars and one TO star identified by the four-dimensional approach. $^{b}$ includes four extra TO stars identified by the four-dimensional approach.}
\end{table}

\section{Members selection} \label{sec:members}
Following a similar approach to \cite{Ordenes2026}, we conducted a search over a 5-dimensional parameter space to identify potential new members of CL001. To do so, we considered our measured astrometric and kinematic data of all stars within a square region of $\approx 5.4' \times 5.4'$, determined by the detector in which CL001 lies. Particularly, we searched their equatorial coordinates, $\alpha$ and $\delta$, their PMs, $\mu_{\alpha*}$ and $\mu_{\delta}$, and the color $J-K_{\rm s}$. After normalizing them using a robust scaler, these five parameters were introduced in an HDBSCAN clustering algorithm~\citep{Campello2013, McInnes2017}. These procedures yield two HDBSCAN clusters: one associated with CL001 and the other with field stars.

Since the clustering algorithm did not account for uncertainties, we applied a Markov-Chain Monte Carlo (MCMC) approach to our measured PMs. The PMs have median relative uncertainties of about $25\%$, which can affect our selection by introducing spurious members with larger PM errors. Therefore, using a Monte Carlo simulation, we drew $50$ PM populations for each star, yielding $50$ probability measurements per star. For a given star, the mean of these probabilities $p$ is considered as its final probability value, and its standard deviation $\sigma$ is considered its error. We defined a relative error as $\sigma_{rel}=p/\sigma$, which is very useful to weight the relevance of the PM error for the result.

Then, we defined four membership classes by using several criteria to select only the most probable members of the cluster. Firstly, the core members with the highest $p$, $p-\sigma\geq 0.8$, and $\sigma_{rel} \leq 0.3$. These are the stars with robust results and the most probable members of the cluster. Secondly, high confidence members with $p \geq 0.8$, $p-\sigma\geq 0.6$, and $\sigma_{rel} \leq 0.4$. This group contains stars with lower precision but with a very good probability of being cluster members. Thirdly, the likely members with $p \geq 0.8$, $p-\sigma\geq 0.6$, and $\sigma_{rel} \leq 0.8$. Lastly, a fourth group, called possible members with $p \geq 0.6$, has a lower probability and error precision, but is very relevant for the faint part of the CMD, where the members are around the TO. In the latter group, we wanted to add the stars that could be rejected from the previous selections only because they are fainter sources with more photometric error.

Due to the inclusion of the $J-K_{\rm s}$ color in the HDBSCAN procedure, this five-dimensional parameter space selection yields stars that are mainly clustered around the vertical RGB. This could leave outside the selection stars that belong to the cluster due to the color $J-K_{\rm s}$ inclusion in the first search. In fact, since HB and TO stars have higher effective temperatures, they will have markedly distinct colors from those of RGB stars. They will therefore be hard to identify as cluster members by our defined HDBSCAN procedure. To test this idea and search for possible HB and TO members of CL001, we applied another approach using a four-dimensional parameter space, only including the equatorial coordinates, $\alpha$ and $\delta$, and the PMs, $\mu_{\alpha*}$ and $\mu_{\delta}$. We use these parameters to find HB candidates in the distance section \ref{subsec:distance} and an extra group around the TO for the age section \ref{subsec:age}.

When applying this four-dimensional HDBSCAN selection, we limit contamination from the foreground by retaining only sources bluer than the isochrone whose distance along the extinction vector from the isochrone is less than 2$\sigma$. Only two HB (high confidence) and five TO stars (one high confidence and four possible) survive to this selection. Using the previous selections, we obtained a final list of $155$ members of VVV CL001. Their location in the CMDs is in the left and middle panels of Fig.~\ref{fig:members}, where the members are color-coded by class, from dark green (core) to light green (possible). Field stars are shown in gray. The blue crosses are the other candidates pre-selected by the four-dimensional approach. The vector point diagram (VPD) of the members and the field stars is shown in the upper right panel. Stars selected as members clearly show a distinct movement pattern compared to the bulk of the field. The bottom right panel shows the sky locations of the selected cluster members. In all panels, black squares indicate cluster members that were spectroscopically confirmed by OC22. The majority of these stars are also marked as core or high-confidence members by our method, indicating excellent agreement with our results. Some of them are recovered with the four-dimensional approach and were avoided since their \J \ magnitude is saturated or blended and for that reason were excluded from the analysis. Overall, a significant number of new members are included in this methodology, significantly increasing the cluster membership. All identified cluster members and their membership classes are summarized in Table~\ref{tab:membership}.

\begin{figure}
\centering
\includegraphics[width=\hsize]{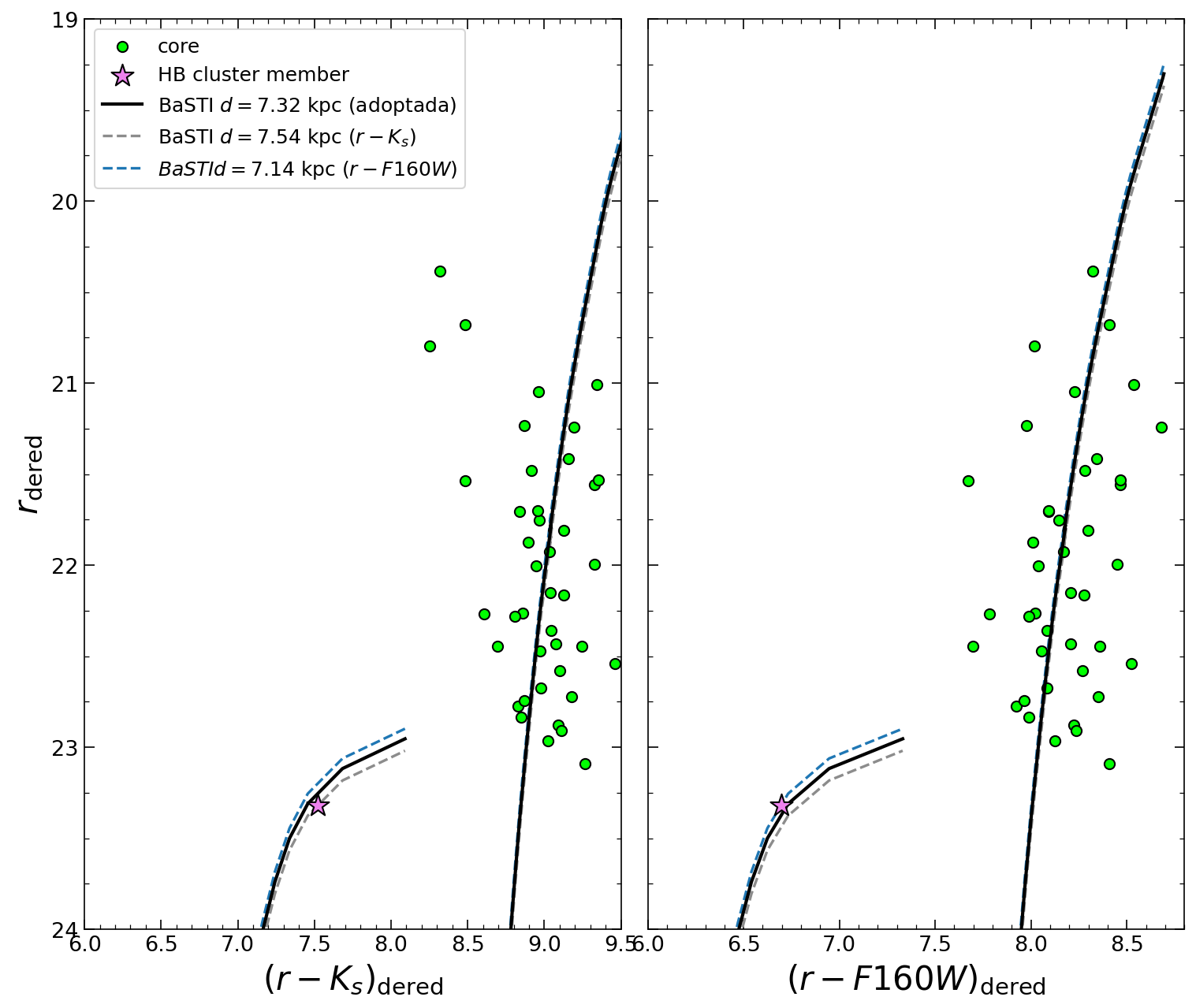}
  \caption{CMD combining optical and NIR regimes. \textit{Left panel}: CMD combining the DECam $r$ and the FourStar $K_{\rm s}$ bands. The black line is a BaSTI isochrone of $13$ Gyr with a metallicity of $-2.2$ dex. The green points are core members in this region. The violet star is the confirmed HB star identified by PMs using photometry across all bands. \textit{Right Panel:} CMD using the DECam $r$ and the HST F160W bands.}
     \label{fig:CMD_HB}
\end{figure}

\begin{table*}
\centering
\scriptsize
\caption{Members of VVV-CL001. Identifiers follow our own numbering
(\texttt{VVV-CL001-M}$nnn$, sorted by right ascension). \J \ and \Ks \ magnitudes from FourStar, F125W and F160W from HST and $r$ from DECam. PMs from the analysis of this work. Classes are those of the HDBSCAN run. The full table is available at the CDS.}
\label{tab:members}
\begin{tabular}{lrrrrrrrrrrrl}
\toprule
ID & $\alpha$ & $\delta$ & $l$ & $b$ & \J \ & \Ks \ & $F125W$ & $F160W$ & $r$ &
$\mu_{\alpha}\cos\delta$ & $\mu_{\delta}$ & Class \\
 & (deg) & (deg) & (deg) & (deg) & \multicolumn{5}{c}{(mag)} &
\multicolumn{2}{c}{(mas\,yr$^{-1}$)} & \\
\midrule
M001 & 268.65356 & $-$24.00664 & 5.2636 & $+$0.8024 & 18.112 & 16.461 &   &   &   & $-$1.609 & $-$2.266 & possible \\
M002 & 268.65906 & $-$24.02083 & 5.2539 & $+$0.7909 & 17.819 & 16.054 &   &   &   & $-$1.964 & $-$1.755 & high\_conf \\
M003 & 268.66077 & $-$24.01103 & 5.2631 & $+$0.7945 & 18.631 & 16.882 & 18.472 & 17.417 &   & $-$1.390 & $-$2.692 & high\_conf \\
M004 & 268.66152 & $-$24.02237 & 5.2537 & $+$0.7881 & 18.370 & 16.662 & 18.366 & 17.349 &   & $-$0.683 & $-$2.188 & possible \\
M005 & 268.66208 & $-$24.01681 & 5.2587 & $+$0.7905 & 15.336 & 13.524 & 15.508 & 14.430 & 22.550 & $-$2.803 & $-$2.208 & high\_conf \\
M006 & 268.66279 & $-$24.01703 & 5.2589 & $+$0.7898 & 18.274 & 16.598 & 18.460 & 17.391 &   & $-$2.333 & $-$1.568 & core \\
M007 & 268.66345 & $-$24.01457 & 5.2613 & $+$0.7906 & 15.056 & 13.206 & 15.221 & 14.097 & 22.496 & $-$3.038 & $-$1.573 & core \\
M008 & 268.66434 & $-$24.01446 & 5.2618 & $+$0.7899 & 19.906 & 18.189 &   &   &   & $-$2.032 & $-$3.120 & possible \\
M009 & 268.66473 & $-$24.02184 & 5.2556 & $+$0.7859 & 15.500 & 13.650 & 15.710 & 14.595 & 23.012 & $-$2.786 & $-$1.431 & possible \\
M010 & 268.66480 & $-$24.01622 & 5.2605 & $+$0.7887 & 18.075 & 16.429 & 18.741 & 17.677 &   & $-$1.507 & $-$1.980 & likely \\
\multicolumn{13}{c}{\dots} \\
\bottomrule
\end{tabular}
\end{table*}

\begin{figure*}
\sidecaption
\includegraphics[width=12cm]{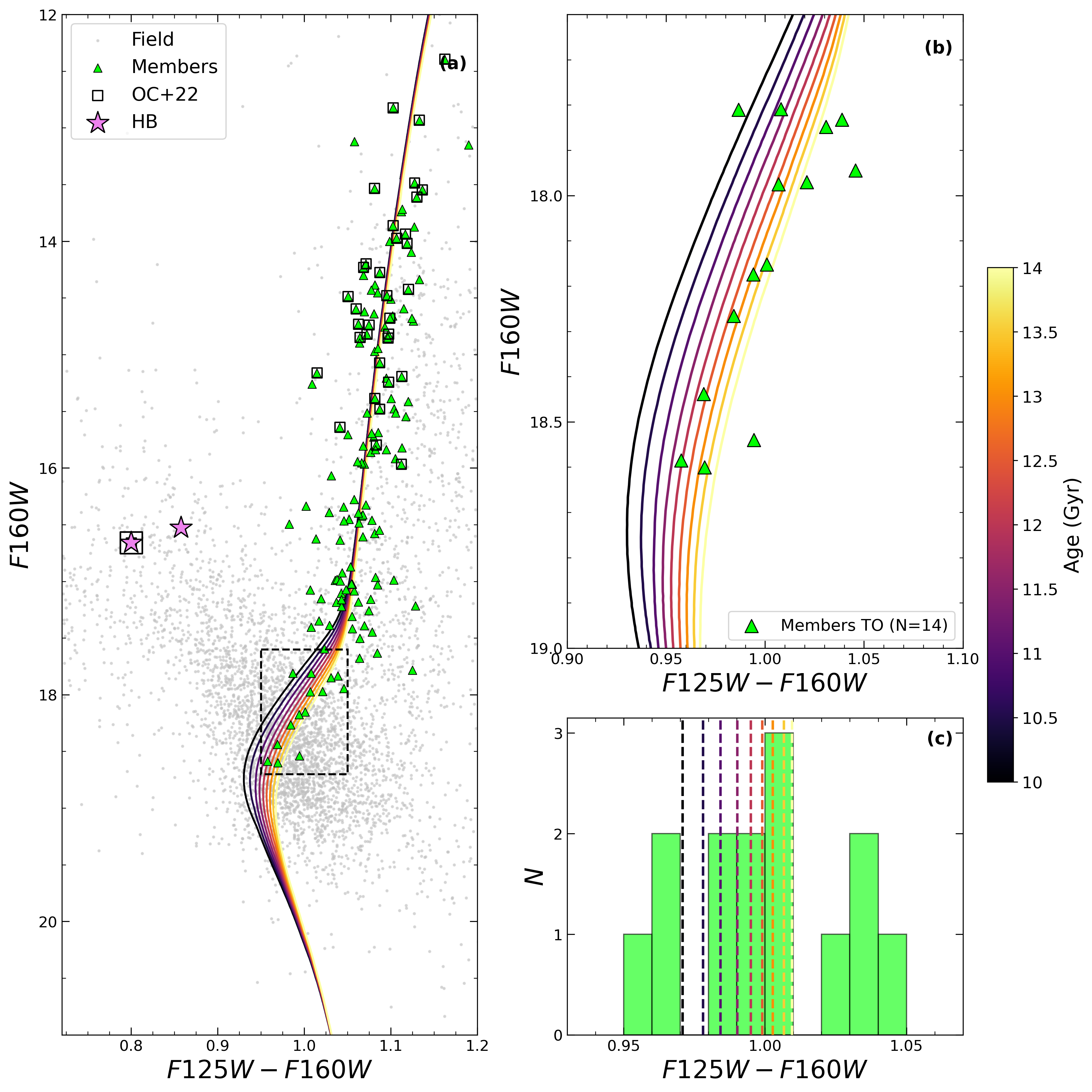}
  \caption{Isochrone fitting around the TO region to calculate the cluster age. \textit{(a):} The complete CMD using the HST filters. Green triangles are all the selected cluster members. The two violet stars are the HB members. Black empty boxes are the OC22 confirmed members, and gray points are field stars. The BaSTI isochrones are color-coded by age. \textit{(b):} Same CMD with a zoom around the TO region. \textit{(c):} The histogram showing the distribution of the TO cluster members. The dashed lines show the isochrone positions in the $F125W-F160W$ color.}
     \label{fig:CMD_TO_basti}
\end{figure*}

\section{Distance and age estimation} \label{sec:distage}
\subsection{Distance} \label{subsec:distance}
We searched for RR Lyrae variable stars around the cluster using the currently available catalogs \citep{Soszynski2019,Clementini2023,Zoccali2024} and, using the available positions, distances, PMs, and metallicities, we attempted to identify cluster members. None of the RR Lyrae stars identified so far around VVV CL001 appears to be a member of the cluster. This is likely due to the cluster's low mass, which leads to a small HB population and a reduced likelihood of finding RR Lyrae stars in the instability strip. 

Due to the lack of RR Lyrae, in order to measure the distance to VVV CL001 precisely, we use the isochrone-fitting technique around the HB in the CMD of the cluster, based on the BaSTI isochrones~\citep{Hidalgo2018,Pietrinferni2021}. For the isochrone, we assume a pure old population with an age of 12 Gyr, given that the cluster is most likely very old and that there is no significant difference in the HB between 10 and 13 Gyrs. For the metallicity, we used a value of $\rm [Fe/H]=-2.2\ dex$, which is the mean from previous studies (FT21, OC22, HM26). 

We created two CMDs based on the cluster members confirmed by our method. One CMD has the $r$ and \Ks \ bands from DECam and FourStar, respectively, and the other has the $r$ and $F160W$ bands from DECam and HST, respectively. Both CMDs are in Fig.~\ref{fig:CMD_HB}.

To select the cluster members in the HB region, we searched for stars near the color of the HB isochrone track and checked whether they remained there, combining the filters: $r$, \Ks \, \J \, $F160W$, $F125W$. If some of the candidates change the color position, they are discarded as members of the HB. Finally, only 2 stars meet these criteria, and only one has $r$ band photometry. Furthermore, the other star was previously confirmed spectroscopically in OC22, and its spectrum shows features characteristic of a hot star, such as the Paschen lines at 8467 and 8862 \AA. More details on this star are given in Appendix ~\ref{app:HBmember}. Then, we adjust the HB isochrones for both CMDs, considering the position of the HB cluster member (violet star) and minimizing the distances of this star to the isochrone track. 

The lack of HB members in this CMD, compared with the suitable values for the R-parameter for GCs (around $R_{HB/RGB} \sim 0.9$) as noted in \cite{Zoccali2000} at the lower limit, can be explained mainly by the high extinction in the field and the cluster metallicity. VVV CL001 is a very metal-poor cluster, which likely results in a more pronounced blue HB tail. Consequently, only the brighter stars were detected in this CMD, leaving many potential HB stars faint and below the DECam $r$ band magnitude limit along the vertical blue tail.

\begin{figure*}
\sidecaption
\includegraphics[width=12cm]{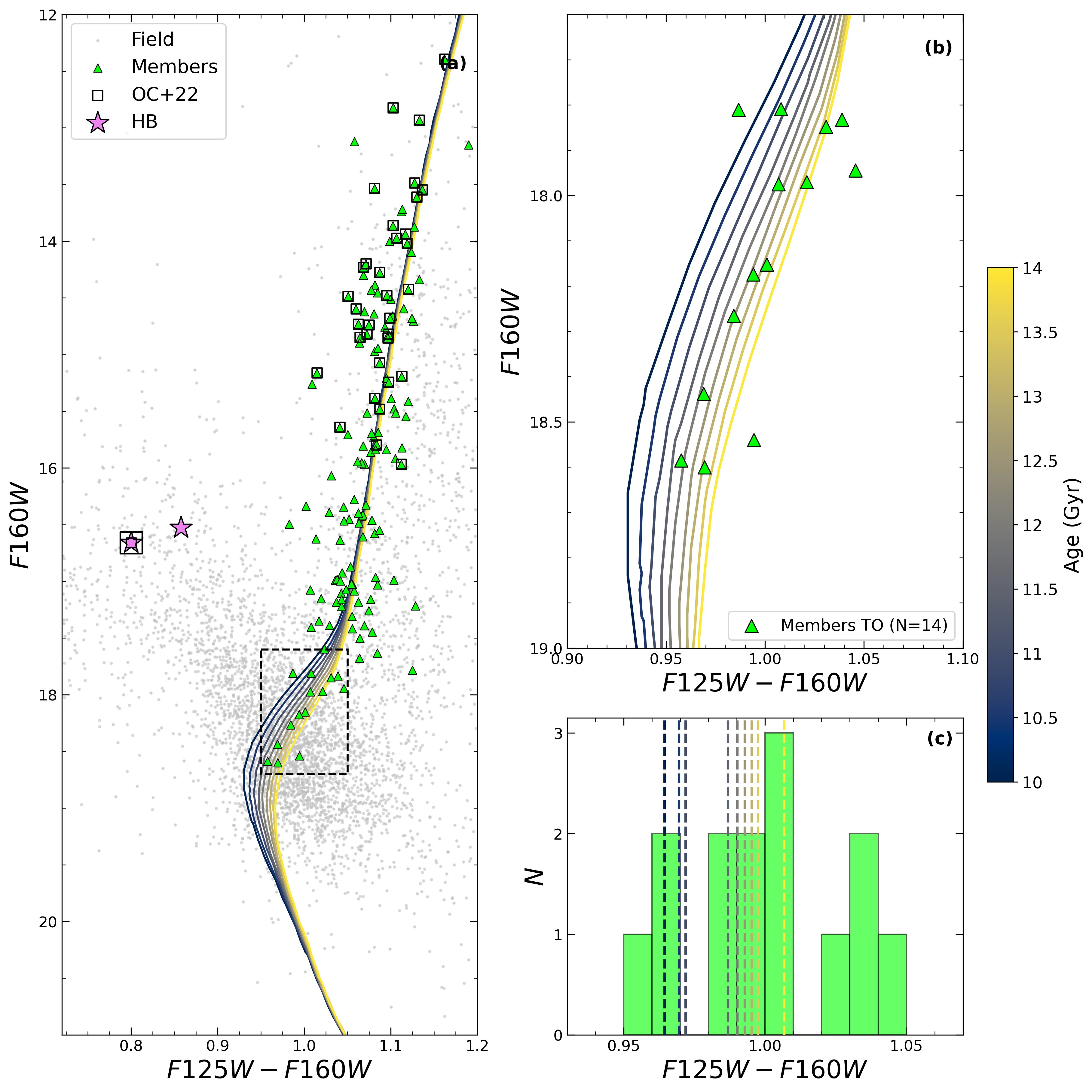}
  \caption{Isochrone fitting around the TO region to calculate the cluster age. \textit{(a):} The complete CMD using the HST filters. Green triangles are all the selected cluster members. The two violet stars are the HB members. Black empty boxes are the OC22 confirmed members and gray points are field stars. The PARSEC isochrones are color coded by age. \textit{(b):} Same CMD with a zoom around the TO region. \textit{(c):} The histogram showing the distribution of the TO cluster members. The dashed lines show the isochrone position in the color $F125W-F160W$.}
     \label{fig:CMD_TO_parsec}
\end{figure*}

To calculate the distance, we used the standard distance equation
\begin{equation}
    (m_\lambda-M_{\lambda})-A_{\lambda}= 5 \log{d}-5,
\end{equation}
where $(m_\lambda-M_\lambda)$ is the distance modulus, and $A_\lambda$ is the extinction coefficient which can be calculated as
\begin{equation}
    A_{\lambda}=R_\lambda\times E(\lambda_1-\lambda_2),
\end{equation}
where $R_{\lambda}$ is the extinction ratio, which for this work is \cite{Nishiyama2009}.

VVV CL001 is located in a very patchy region where the extinction varies in a range between $E(J-{K_S})=1-1.4$ mag, so the reddening has to be used as a free parameter. With the isochrone fitting for the $r$ and \Ks \ CMD, we obtained a distance modulus of $(r-M_{r})=22.58$ mag and a reddening of  $E(r-{K_s})=7.057$ mag. This value is consistent with the one obtained in NIR bands $E(J-K_s)=1.267$ mag and $E(F125W-F160W)=0.74$ based on the RGB members using the \cite{Nishiyama2009} extinction law. With these values, the cluster distance is $ d=7.54\pm0.26$ kpc. 

Moreover, for the $r$ and $F160W$ CMD, we obtained a distance modulus of $(r-M_{r})=22.46$ mag and a reddening of  $E(r-{F160W})=6.37$ mag. With these values, the cluster distance is $d=7.14\pm0.45$ kpc. Thus, the mean distance is $d = 7.34\pm0.46$ kpc, consistent with previous studies and placing VVV CL001 in the inner region of the bulge. The error accounts for the different distances obtained and the reddening-induced error, the main source of uncertainty, which is the quadratic sum of the statistical error of 0.45 kpc and the systematic error of 0.2 kpc. This is the smallest distance error reported in the literature so far; however, in the next section, it is clear that distance error remains the largest source of uncertainty for the age. New, deeper observations of the cluster in optical bands that cover the fainter HB cluster members will soon be available to address this problem.

\subsection{Age} \label{subsec:age}
To estimate the cluster age precisely, it is essential to have a TO well defined by the cluster members. This has been a limitation in all previous studies, since VVV CL001 is embedded in a region with patchy extinction, high stellar density, and an intrinsic low luminosity, requiring very deep photometry to resolve these issues. We can reach the TO of the cluster using deep HST wide-filter photometry, which is optimized for deep photometry in crowded fields. We used the F125W and F160W filters in the near-infrared to construct the CMD and used different isochrones, varying only the age, to determine the final value. These two near-IR filters overlap, so they suffer less from differential reddening, allowing for the recovery of features such as the TO with more precision.

We first used the BaSTI isochrones with a fixed metallicity of -2.2 dex, spanning 10 to 14 Gyr in 0.5 Gyr steps. This selection is based on previous results that put VVV CL001 as an old globular cluster (FT21, HM26). The distance was fixed at 7.34 kpc, and the reddening at $\rm E(F125W-F160W)=0.74$ mag. 

A typical issue with HST-wide filters is the reddening of RGB stars. An offset of at least 0.4 mag exists between the BaSTI isochrone and the current data location, with the isochrone shifted toward the redder part. This problem is usually solved by applying a theoretical $\rm T_{eff}$ correction to the isochrones. The $\rm T_{eff}$ correction is based on the fact that lower-temperature stars experience different levels of extinction; thus, it is recalculated as a function of the star's temperature~\citep{Girardi2008,Chen2019}. See Appendix ~\ref{app:CMDisochronescorr} for more details.

Thus, we applied the $\rm T_{eff}$ correction to the BaSTI isochrones. For each point of the isochrone, the extinction is recalculated by the equation:
\begin{equation}
\mathrm E_i=E_{\mathrm{TO}}-k\times S(T_{\mathrm{eff,i}})
\end{equation}
where $S$ is the shape of the black body and is based on the BaSTI column "logTe". The factor $k$ is an empirical parameter calibrated by iterating until the median of the residuals around the RGB is zero. In this case, for the BaSTI HST isochrone, the value is $k=6.24$. A similar correction for HST imaging in the GC context was used in previous studies~\citep{Cohen2015,Hidalgo2018,Massari2023}. Figure \ref{fig:CMD_TO_basti} shows the CMD with corrected isochrones of different ages fitted to the cluster members. The zoomed region shows the best isochrones around the TO zone. We found that the best-fit isochrones are those at 13-13.5 Gyr, and using $\chi^2$ and the rms, we obtain an age of 13.47 Gyr.  

Furthermore, we used the PARSEC isochrones~\citep{Bressan2012,Tang2014} to check the result of the BaSTI ones. We repeated the fixed metallicity of -2.2 dex, spanning 10 to 14 Gyr in 0.5 Gyr steps. Additionally, the PARSEC isochrones allow us to apply the extinction factor $A_V$ directly, and we then added the calculated value $A_V = 9.23$. So we obtained a directly dereddened PARSEC isochrone and used it to calculate the age. Figure~\ref{fig:CMD_TO_parsec} shows the CMD with this set of isochrones fitted to the cluster members. We found that the best-fit isochrones have ages around 13-13.5 Gyr, and using $\chi^2$ and the rms, we calculate an age of 13.39 Gyr. This result confirms the previous one, and by combining both values, we estimate an age of $13.4 \pm 0.4$ Gyr for the VVV CL001 cluster. A systematic error of around 1 Gyr must be added, accounting for distance-error propagation. Even with the very deep HST photometry, the TO is not completely covered in the presented CMDs. A follow-up study reaching more magnitudes below the TO is necessary to solidify this result.  

The age obtained for the cluster is relevant considering the age of the Universe, which is in the range 13.6-13.8 Gyr. Hence, VVV CL001 is one of the most metal-poor GCs in the bulge and, given its age, is possibly the oldest GC in the entire Galaxy. An important point is that there is a limit on the age (age of the Universe), indicating that the cluster's distance can be farther but not much closer than the one calculated in this work.


We compare the age of VVV CL001 with that of other bulge GCs. For instance, there are other candidates for the oldest GC in the MW. The bulge GC HP1 has a metallicity of -1.06 dex and an age of 12.9 Gyr \citep{Kerber2019}. NGC 6256 is another bulge GC with a metallicity of -1.02 dex and an age of 13 Gyr \citep{Cadelano2020}. NGC 6558 is also a bulge GC with a metallicity of -1.17 dex and an age of 13.2 Gyr \citep{Souza2024}. For these clusters, the age is very close to that of VVV CL001, but the metallicity is higher, which could favor VVV CL001 as older, since it may have formed in an earlier, more pristine gas cloud \citep{Geisler2025}. Overall, these are the oldest GCs in the Galaxy and are located in the Galactic bulge, where they all currently form part of the metal-poor spheroid.

\begin{figure*}
\centering
\includegraphics[width=16cm]{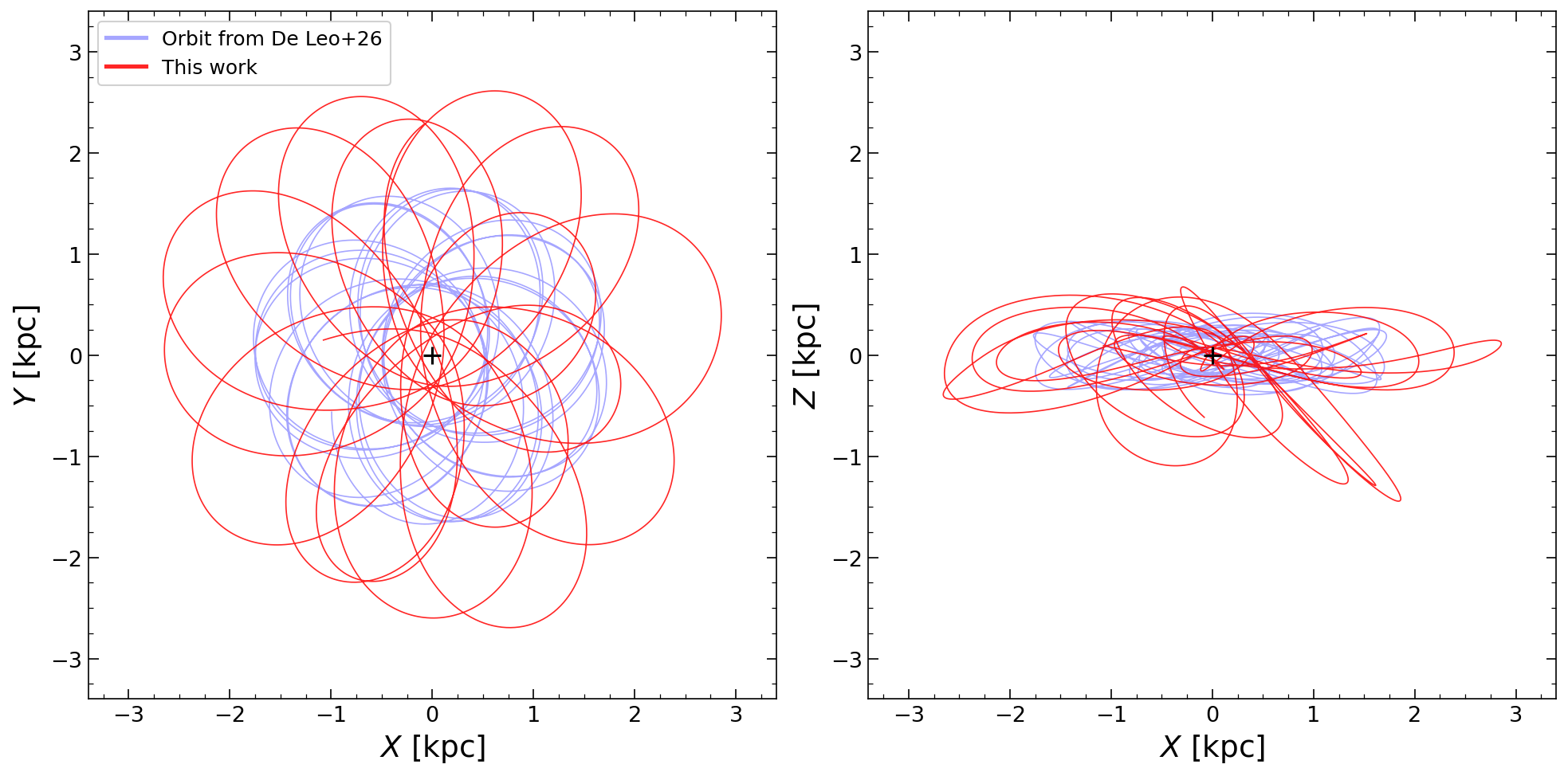}
  \caption{The orbit of the VVV CL001 cluster. The red line represents the new orbit based on the parameters used in this work. The blue line is the old orbit from \cite{DeLeo2026}, using a distance of 8.23 kpc.}
     \label{fig:orbit}
\end{figure*}

\section{Orbital integration} \label{sec:orbit}
To derive the orbit of VVV CL001, we followed the same approach presented in \cite{Olivares2024,Olivares2026}. Here, we present a brief summary of the method. 

We used the orbital code \texttt{OrbIT}~\citep{DeLeo2026}, which has the advantage of being customizable and including a time-varying potential that includes the rotating bar of the MW, very influential in the inner part of the Galaxy. The model of the MW gravitational potential used in our orbital integrator includes a Navarro-Frenk-White dark matter halo~\citep{Navarro1996}, two stellar and two gaseous disks with the profile by \cite{Miyamoto1975}, a rotating bar~\citep{Long1992}, and a spherical bulge component modeled as a Hernquist profile~\citep{Hernquist1990}. The details of the masses and all other parameters for each component of the potential can be found in \cite{DeLeo2026}.

To derive the orbital parameters, we converted our best estimates of the coordinates, RVs, PMs, and distances into the Cartesian galactocentric frame using the Astropy modules~\citep{Astropy2013,Astropy2018}. We used a distance of the Sun to the Galactic center of 8.27 kpc~\citep{GRAVITY2021} and the solar velocity vector $\rm (U,V,W)=(11.1,12.24,7.25)\ km \ s^{-1}$~\citep{Schonrich2010}. The velocity of the Local Standard of Rest is assumed to be $\rm V_{LSR} = 232.8 \ km \ s^{-1}$~\citep{McMillan2017}. For the cluster, the orbit was evolved backward in time for 5 Gyr inside the MW potential. The cluster orbital parameters are summarized in Table~\ref{tab:params}.

Figure~\ref{fig:orbit} in both panels shows the new orbit derived for the VVV CL001 cluster (red) using the parameters derived in this work. We compared this orbit with an old one (blue) presented in \cite{DeLeo2026}, which had a distance of 8.23 kpc based on OC22. With the new distance, the VVV CL001 orbit becomes more radial; however, it remains within the Galactic bulge. This new orbit is more related to the spheroid of the bulge~\citep{Zoccali2018,Lim2021}, and thus, to the metal-poor population of the bulge. Considering the energy and angular momentum of the orbit, it appears to be a bulge member but is close to the low-energy group of clusters (as characterized in \cite{DeLeo2026}). However, when all orbital parameters are taken into account, the cluster appears to be a member of the \textit{in situ} bulge family. \cite{DeLeo2026}. Hence, we conclude that VVV CL001 is an ancient and most likely the most metal-poor bulge GC. 

We can compare our orbital results with some of the literature. FT21 found that VVV CL001 has a more likely halo orbit due to the high RV and the distance obtaining large apocenter radius and $z_{max}$, but they did not have a precise distance at that time, so they assumed different distance values. More recently, HM26 with a new value for the distance, found that due to the apocenter radius of around 5 kpc and the $z_{max}>1$ kpc, and considering the low metallicity, the cluster was formed in the very early disk before the formation of the bar. In our case, all the orbital parameters place the cluster in the bulge; for that reason, we associated the cluster with the metal-poor bulge spheroidal component \citep{ArdernArentsen2024}, which is also older than the bar.

\begin{table}
\centering
\caption{Final parameters of VVV-CL001.}
\label{tab:params}
\begin{tabular}{llll}
\toprule
Parameter & Value & Unit \\
\midrule
$\alpha$ (J2000), $\delta$ (J2000)      & 17:54:42.4, $-$24:00:55.4   &       \\
$l,\ b$               & 5.2668, $+$0.7796            & deg    \\
$RV_{\rm helio}$      & $-324.9 \pm 0.8$             & km\,s$^{-1}$ \\
$\sigma_{RV}$         & 7.2                          & km\,s$^{-1}$  \\
$[\mathrm{Fe/H}]$      & $-2.2 \pm 0.2$                      & dex     \\
$\mu_{\alpha}\cos\delta$ & $-3.309 \pm 0.047$        & mas\,yr$^{-1}$  \\
$\mu_{\delta}$        & $-1.775 \pm 0.036$           & mas\,yr$^{-1}$ \\
$E(J-K_s)$            & $1.267 \pm 0.004$            & mag    \\
$E(F125W-F160W)$      & $0.743 \pm 0.002$          & mag     \\
$E(r-K_s)$            & $7.057 \pm 0.043$            & mag     \\
$(m-M)_0$             & 14.323                       & mag     \\
$d$                   & $7.34\pm0.46$      & kpc    \\
Age                   & $13.4 \pm 0.4$    & Gyr     \\
$r_{\rm apo}$         & $2.24 \pm 0.28$              & kpc   \\
$r_{\rm peri}$        & $0.65 \pm 0.40$              & kpc     \\
$z_{\rm max}$         & 2.16                        & kpc     \\
$e$                   & $0.563 \pm 0.241$            &         \\
circularity           & 0.276                        &        \\
$E_{tot}$             & $-1.98\cdot 10^5$            &km$^{2}$\,s$^{-2}$        \\
$L_z$                 & 111.49                       &kpc\,km\,s$^{-1}$        \\
\bottomrule
\end{tabular}
\end{table}

\section{GCs as chronometers of the MW assembly} \label{sec:discussion}

The oldest Galactic GCs are among the most ancient objects whose ages can be measured accurately and precisely, tracing the Milky Way assembly.
With ages $\gtrsim 10$~Gyr, GCs define a marked age--metallicity relation (AMR), with the oldest clusters also being the most metal-poor \citep{Ying2025}, 
having formed within the first $\sim 1$~Gyr \citep[$z \gtrsim 6$;][]{MartinFranch2009, VandenBerg2013, Valcin2025}.
Combining the AMR with the cluster dynamical properties, it is possible to infer the progenitor systems (or accretion events), 
since each of them defines its own tight AMR sequence, a classic signature of accretion from disrupted dwarf galaxies \citep{Massari2025, AguadoAgelet2026}.
This analysis is now precise enough to isolate individual accretion events, e.g., Gaia-Sausage/Enceladus, Sagittarius, Sequoia, the Helmi streams, 
and the ``Low-energy-Kraken-Heracles'' (LKH) merger and to separate them from the {\it in situ} or proto-MW population \citep{Massari2019, Massari2026}.
In this way, the metal-poor, low-energy, low-angular-momentum clusters confined to the inner few kiloparsecs are among the best stellar tracers of the early MW assembly.
In this inner region, a centrally concentrated, metal-poor (${\rm [Fe/H] \lesssim -1.5}$), 
kinematically hot and massive stellar population containing $\gtrsim 5\times10^{7} M_{\odot}$ called \textit{Aurora} has been characterized, 
implying a formation epoch of $z \gtrsim 5$ \citep{Rix2022, ArdenArentsen2024}.
Under stricter chemo-dynamical constraints, {\it Aurora} was defined as the population with distinct [N/Fe] and [Al/Fe] content, 
and limited to the bona fide {\it in situ} population, formed before the Galactic disk spin-up \citep{Belokurov2022}.
This feature was independently recovered in the GC population \citep{Belokurov2024}, 
placing in the proto-MW category clusters more metal-rich than VVV~CL001 ($-1.3 \lesssim {\rm [Fe/H]} \lesssim -1.0$), with ages consistently around $\sim 12.5$~Gyr.
Although VVV~CL001 lies roughly $1$~dex below in metallicity (${\rm [Fe/H] \approx -2.2}$), its older age ($\sim 13.4$~Gyr) compensates for this offset, 
naturally extending the AMR of the moderately metal-poor bulge clusters defined by \citet[][see their Fig.~10]{Souza2024}.

Until recently, the environment in which the first GCs formed could only be inferred indirectly, from the fossil record of the surviving systems.
JWST has modified this paradigm, resolving compact, dense stellar structures inside strongly lensed 
galaxies from $z \sim 1$ out to $z \sim 10$ \citep{Vanzella2017, Claeyssens2023, Messa2024, Adamo2024}.
These cluster candidates share several properties: very high stellar densities, young ages ($\lesssim 10$--50~Myr at the epoch of observation), 
low but non-negligible metallicities, and clumpy, turbulent, gas-rich hosts.
This is precisely the regime in which the cluster formation efficiency is predicted 
to rise steeply with the surface density of star formation \citep{Kruijssen2012, Kruijssen2015, Pfeffer2018}.

VVV~CL001 most likely belongs to the surviving end of this primordial GC population 
\citep[see the insightful discussion on][]{Geisler2023, Geisler2025}, which raises two related issues: 
i) in which region of the proto-MW was the cluster formed? was it born in the innermost region, or did it migrate there at a later time? 
and ii) how did it survive a Hubble time in the inner Galaxy where strong dynamical processes take place? The two are connected through the GC mass.
A cluster of $\sim 10^{5} M_{\odot}$ cannot sink to the inner kiloparsec under its own dynamical friction \citep{Pagnini2023}, 
so an \textit{accreted} origin would require that VVV~CL001 was carried inwards while still bound to a massive host, 
implicating an additional very early LKH-type merger event rather than a normal dwarf galaxy accretion event (or even within the LKH accretion event).
An {\it in-situ} origin, in turn, does not imply that the present orbit is primordial, 
since resonant trapping by the growing bar can spin up and flatten initially pressure-supported inner-halo orbits \citep{Dillamore2024}.
In both cases, survival at such low pericentres, where tidal shocking disrupts most clusters \citep{Fall2001, Li2019}, 
implies an initial mass of the order of $10^{6} M_{\odot}$ \citep[][see their Fig.~7]{Baumgardt2019}.
In the absence of a comprehensive chemical characterization, VVV~CL001 would then probe the high-mass end of the GC initial mass function \citep{Kruijssen2015, ReinaCampos2019}.

\section{Conclusions} \label{sec:conclusions}

We present here new distance, age and orbit estimations of the bulge GC VVV CL001, a low-mass, metal-poor GC in a very crowded and highly extincted region of the sky. We used the multi-epoch NIR observations from FourStar to determine PMs of the stars in the cluster field of view. Using the HDBSCAN algorithm together with an MCMC approach, we determine the stellar members of the cluster. We found a significant number of new cluster members, including those in the HB and TO evolutionary stages, bringing the total to 155 stars. 

Using the optical DECam $r$ band, the NIR Fourstar \Ks \ band, and the NIR HST $F160W$ band, we confirmed two members in the HB of the cluster. Moreover, one has a spectrum from OC22 that shows features of a hot star consistent with being on the HB. We use the BaSTI isochrones for the HB and RGB members to determine the distance to the cluster, obtaining $d=7.34\pm0.46$ kpc, in agreement with the most recent reported value by HM26 of $d=7.1 \pm 1$ kpc, placing the cluster within the Galactic bulge. It is also consistent with the FT21 determination within their rather large error  $d=8.22 \pm 2$ kpc. Our result also has the smallest distance error reported in the literature to date. Furthermore, we will receive new optical observations to improve this result in future work.

We estimate the age of the cluster using the HST $F125W$ and $F160W$ filters due to their depth and less sensitivity to reddening. Several TO cluster members were selected to calculate ages using the isochrone-fitting technique with the BaSTI and PARSEC isochrones independently. Both methods estimate the cluster age at $13.4\pm 0.4$ Gyr, suggesting that VVV CL001 could be the oldest GC in the bulge and perhaps the Galaxy, given its metallicity of $-2.2$ dex. This age contrasts with previous estimates from FT21 and HM26, which are around 12 Gyr; however, they did not reach the TO required to robustly estimate the age, so the uncertainty in those studies is larger. Using our method and the distance reported in HM26, the age of VVV CL001 is 13.7 Gyr, very close to the age of the Universe.

We derived and updated the cluster orbit using the new parameters with the \texttt{OrbIT} code. We found that VVV CL001 is a bona fide bulge cluster on a radial orbit, consistent with the metal-poor bulge spheroid. Confirming VVV CL001 as an \textit{in situ} GC is crucial, as it has already been spectroscopically confirmed as the most metal-poor GC in the bulge so far.

We are preparing a high-resolution spectroscopic follow-up of the cluster to solidify the results of this paper by studying several chemical elements. By studying these abundances, we will be able to confirm an external origin or identify other correlations~\citep{Belokurov2024}. If this is an \textit{in situ} GC of this age, the age of the Galaxy, and perhaps the Universe, must be reconsidered in light of the presence of these extremely ancient fossils. 

\section*{Data availability}
The full Table~\ref{tab:members} is only available in electronic form at the CDS via anonymous ftp to cdsarc.u-strasbg.fr (130.79.128.5) or via http://cdsweb.u-strasbg.fr/cgi-bin/qcat?J/A+A/.

\begin{acknowledgements}
      J.O.C. acknowledges financial support from the project "Partecipazione LSST - Large Synoptic Survey Telescope" (Ob.Fu. 1.05.03.06; local PI: M. Brescia). This work has been supported by the ANID BASAL Center for Astrophysics and Associated Technologies (CATA) through grant FB210003 AIM23-0001. A.V.N. acknowledges support from the National Agency for Research and Development (ANID) Scholarship Program Doctorado Nacional 2020–21201226. C.S.G. acknowledges support from the National Agency for Research and Development (ANID) Doctorado Nacional grant 2025-21251677. C.O.H. acknowledges Agencia Nacional de Investigación y Desarrollo (ANID) through FONDECYT postdoctoral grant 3260854. B.A.T. acknowledges support from the National Agency for Research and Development (ANID) Doctorado Nacional grant 21231305/2023. F.G. acknowledges support from the Research fellow Gemini-ANID 2025 - 32RF250005 project, and the international Gemini Observatory, a program of NSF NOIRLab, which is managed by the Association of Universities for Research in Astronomy (AURA) under a cooperative agreement with the U.S. National Science Foundation, on behalf of the Gemini partnership of Argentina, Brazil, Canada, Chile, the Republic of Korea, and the United States of America. M.Z. acknowledges support from FONDECYT Regular grant No. 1230731. S.V. gratefully acknowledges the support provided by the ANID BASAL project FB210003. D.G. acknowledges financial support from the Vicerrectoría de Investigación y Postgrado de la Universidad de La Serena.

      It also made use of NASA’s Astrophysics Data System and of the VizieR catalog access tool, CDS, Strasbourg, France \citep{simbad}.  The original description of the VizieR service was published in \citep{vizier}. Finally, we acknowledge the use of the following publicly available softwares: DAOPHOT \citep{stetson1987}, DOLPHOT \citep{Dolphin2016}, TOPCAT \citep{topcat}, pandas \citep{pandas}, IPython \citep{ipython}, numpy \citep{numpy}, matplotlib \citep{matplotlib}, Astropy, a community developed core Python package for Astronomy \citep{astropy1,astropy2} and Aladin sky atlas \citep{aladin1, aladin2}. 
\end{acknowledgements}

\bibliographystyle{aa} 
\bibliography{CL001} 

@ARTICLE{McInnes2017,
       author = {{McInnes}, Leland and {Healy}, John and {Astels}, Steve},
        title = "{hdbscan: Hierarchical density based clustering}",
      journal = {The Journal of Open Source Software},
         year = 2017,
        month = mar,
       volume = {2},
       number = {11},
          eid = {205},
        pages = {205},
          doi = {10.21105/joss.00205},
       adsurl = {https://ui.adsabs.harvard.edu/abs/2017JOSS....2..205M}
}

@InProceedings{Campello2013,
    author="Campello, Ricardo J. G. B.
    and Moulavi, Davoud
    and Sander, Joerg",
    editor="Pei, Jian
    and Tseng, Vincent S.
    and Cao, Longbing
    and Motoda, Hiroshi
    and Xu, Guandong",
    title="Density-Based Clustering Based on Hierarchical Density Estimates",
    booktitle="Advances in Knowledge Discovery and Data Mining",
    year="2013",
    publisher="Springer Berlin Heidelberg",
    address="Berlin, Heidelberg",
    pages="160--172",
    isbn="978-3-642-37456-2"
}

@ARTICLE{Haro-Moya+2025,
       author = {{Haro Moya}, W. and {Moni Bidin}, C. and {Parisi}, M.~C. and {Geisler}, D. and {Bla{\~n}a}, M. and {Villanova}, S. and {Mauro}, F. and {Chen{\'e}}, A.-N. and {Cohen}, R.~E. and {Ram{\'\i}rez Alegr{\'\i}a}, S. and {Contreras Ramos}, R. and {Zoccali}, M. and {Mu{\~n}oz}, C. and {Dias}, B.},
        title = "{Infrared photometry and CaT spectroscopy of the most metal-poor in-situ globular cluster VVV-CL001}",
      journal = {arXiv e-prints},
         year = 2025,
        month = nov,
          eid = {arXiv:2511.13161},
        pages = {arXiv:2511.13161},
          doi = {10.48550/arXiv.2511.13161},
archivePrefix = {arXiv},
       eprint = {2511.13161},
 primaryClass = {astro-ph.GA},
       adsurl = {https://ui.adsabs.harvard.edu/abs/2025arXiv251113161H}
}

@ARTICLE{Persson+2013,
       author = {{Persson}, S.~E. and {Murphy}, D.~C. and {Smee}, S. and {Birk}, C. and {Monson}, A.~J. and {Uomoto}, A. and {Koch}, E. and {Shectman}, S. and {Barkhouser}, R. and {Orndorff}, J. and {Hammond}, R. and {Harding}, A. and {Scharfstein}, G. and {Kelson}, D. and {Marshall}, J. and {McCarthy}, P.~J.},
        title = "{FourStar: The Near-Infrared Imager for the 6.5 m Baade Telescope at Las Campanas Observatory}",
      journal = {\pasp},
         year = 2013,
        month = jun,
       volume = {125},
       number = {928},
        pages = {654},
          doi = {10.1086/671164},
       adsurl = {https://ui.adsabs.harvard.edu/abs/2013PASP..125..654P}
}

@ARTICLE{Schirmer2013,
       author = {{Schirmer}, M.},
        title = "{THELI: Convenient Reduction of Optical, Near-infrared, and Mid-infrared Imaging Data}",
      journal = {\apjs},
         year = 2013,
        month = dec,
       volume = {209},
       number = {2},
          eid = {21},
        pages = {21},
          doi = {10.1088/0067-0049/209/2/21},
archivePrefix = {arXiv},
       eprint = {1308.4989},
 primaryClass = {astro-ph.IM},
       adsurl = {https://ui.adsabs.harvard.edu/abs/2013ApJS..209...21S}
}

@ARTICLE{stetson1987,
       author = {{Stetson}, Peter B.},
        title = "{DAOPHOT: A Computer Program for Crowded-Field Stellar Photometry}",
      journal = {\pasp},
         year = 1987,
        month = mar,
       volume = {99},
        pages = {191},
          doi = {10.1086/131977},
       adsurl = {https://ui.adsabs.harvard.edu/abs/1987PASP...99..191S}
}

@ARTICLE{stetson1994,
       author = {{Stetson}, Peter B.},
        title = "{The Center of the Core-Cusp Globular Cluster M15: CFHT and HST Observations, ALLFRAME Reductions}",
      journal = {\pasp},
         year = 1994,
        month = mar,
       volume = {106},
        pages = {250},
          doi = {10.1086/133378},
       adsurl = {https://ui.adsabs.harvard.edu/abs/1994PASP..106..250S}
}

@INPROCEEDINGS{bertin2006,
       author = {{Bertin}, E.},
        title = "{Automatic Astrometric and Photometric Calibration with SCAMP}",
    booktitle = {Astronomical Data Analysis Software and Systems XV},
         year = 2006,
       editor = {{Gabriel}, C. and {Arviset}, C. and {Ponz}, D. and {Enrique}, S.},
       series = {Astronomical Society of the Pacific Conference Series},
       volume = {351},
        month = jul,
        pages = {112},
       adsurl = {https://ui.adsabs.harvard.edu/abs/2006ASPC..351..112B}
}

@ARTICLE{smith2025,
       author = {{Smith}, Leigh C. and {Lucas}, Philip W. and {Koposov}, Sergey E. and {Gonzalez-Fernandez}, Carlos and {Alonso-Garc{\'\i}a}, Javier and {Minniti}, Dante and {Sanders}, Jason L. and {Bedin}, Luigi R. and {Belokurov}, Vasily and {Evans}, N. Wyn and {Hempel}, Maren and {Ivanov}, Valentin D. and {Kurtev}, Radostin G. and {Saito}, Roberto K.},
        title = "{VIRAC2: NIR astrometry and time series photometry for 500M+ stars from the VVV and VVVX surveys}",
      journal = {\mnras},
         year = 2025,
        month = feb,
       volume = {536},
       number = {4},
        pages = {3707-3738},
          doi = {10.1093/mnras/stae2797},
archivePrefix = {arXiv},
       eprint = {2501.06295},
 primaryClass = {astro-ph.GA},
       adsurl = {https://ui.adsabs.harvard.edu/abs/2025MNRAS.536.3707S}
}

@ARTICLE{gaiadr3,
       author = {{Gaia Collaboration} and {Vallenari}, A. and {Brown}, A.~G.~A. and {Prusti}, T. and {de Bruijne}, J.~H.~J. and {Arenou}, F. and {Babusiaux}, C. and {Biermann}, M. and {Creevey}, O.~L. and {Ducourant}, C. and {Evans}, D.~W. and {Eyer}, L. and {Guerra}, R. and {Hutton}, A. and {Jordi}, C. and {Klioner}, S.~A. and {Lammers}, U.~L. and {Lindegren}, L. and {Luri}, X. and {Mignard}, F. and {Panem}, C. and {Pourbaix}, D. and {Randich}, S. and {Sartoretti}, P. and {Soubiran}, C. and {Tanga}, P. and {Walton}, N.~A. and {Bailer-Jones}, C.~A.~L. and {Bastian}, U. and {Drimmel}, R. and {Jansen}, F. and {Katz}, D. and {Lattanzi}, M.~G. and {van Leeuwen}, F. and {Bakker}, J. and {Cacciari}, C. and {Casta{\~n}eda}, J. and {De Angeli}, F. and {Fabricius}, C. and {Fouesneau}, M. and {Fr{\'e}mat}, Y. and {Galluccio}, L. and {Guerrier}, A. and {Heiter}, U. and {Masana}, E. and {Messineo}, R. and {Mowlavi}, N. and {Nicolas}, C. and {Nienartowicz}, K. and {Pailler}, F. and {Panuzzo}, P. and {Riclet}, F. and {Roux}, W. and {Seabroke}, G.~M. and {Sordo}, R. and {Th{\'e}venin}, F. and {Gracia-Abril}, G. and {Portell}, J. and {Teyssier}, D. and {Altmann}, M. and {Andrae}, R. and {Audard}, M. and {Bellas-Velidis}, I. and {Benson}, K. and {Berthier}, J. and {Blomme}, R. and {Burgess}, P.~W. and {Busonero}, D. and {Busso}, G. and {C{\'a}novas}, H. and {Carry}, B. and {Cellino}, A. and {Cheek}, N. and {Clementini}, G. and {Damerdji}, Y. and {Davidson}, M. and {de Teodoro}, P. and {Nu{\~n}ez Campos}, M. and {Delchambre}, L. and {Dell'Oro}, A. and {Esquej}, P. and {Fern{\'a}ndez-Hern{\'a}ndez}, J. and {Fraile}, E. and {Garabato}, D. and {Garc{\'\i}a-Lario}, P. and {Gosset}, E. and {Haigron}, R. and {Halbwachs}, J.-L. and {Hambly}, N.~C. and {Harrison}, D.~L. and {Hern{\'a}ndez}, J. and {Hestroffer}, D. and {Hodgkin}, S.~T. and {Holl}, B. and {Jan{\ss}en}, K. and {Jevardat de Fombelle}, G. and {Jordan}, S. and {Krone-Martins}, A. and {Lanzafame}, A.~C. and {L{\"o}ffler}, W. and {Marchal}, O. and {Marrese}, P.~M. and {Moitinho}, A. and {Muinonen}, K. and {Osborne}, P. and {Pancino}, E. and {Pauwels}, T. and {Recio-Blanco}, A. and {Reyl{\'e}}, C. and {Riello}, M. and {Rimoldini}, L. and {Roegiers}, T. and {Rybizki}, J. and {Sarro}, L.~M. and {Siopis}, C. and {Smith}, M. and {Sozzetti}, A. and {Utrilla}, E. and {van Leeuwen}, M. and {Abbas}, U. and {{\'A}brah{\'a}m}, P. and {Abreu Aramburu}, A. and {Aerts}, C. and {Aguado}, J.~J. and {Ajaj}, M. and {Aldea-Montero}, F. and {Altavilla}, G. and {{\'A}lvarez}, M.~A. and {Alves}, J. and {Anders}, F. and {Anderson}, R.~I. and {Anglada Varela}, E. and {Antoja}, T. and {Baines}, D. and {Baker}, S.~G. and {Balaguer-N{\'u}{\~n}ez}, L. and {Balbinot}, E. and {Balog}, Z. and {Barache}, C. and {Barbato}, D. and {Barros}, M. and {Barstow}, M.~A. and {Bartolom{\'e}}, S. and {Bassilana}, J.-L. and {Bauchet}, N. and {Becciani}, U. and {Bellazzini}, M. and {Berihuete}, A. and {Bernet}, M. and {Bertone}, S. and {Bianchi}, L. and {Binnenfeld}, A. and {Blanco-Cuaresma}, S. and {Blazere}, A. and {Boch}, T. and {Bombrun}, A. and {Bossini}, D. and {Bouquillon}, S. and {Bragaglia}, A. and {Bramante}, L. and {Breedt}, E. and {Bressan}, A. and {Brouillet}, N. and {Brugaletta}, E. and {Bucciarelli}, B. and {Burlacu}, A. and {Butkevich}, A.~G. and {Buzzi}, R. and {Caffau}, E. and {Cancelliere}, R. and {Cantat-Gaudin}, T. and {Carballo}, R. and {Carlucci}, T. and {Carnerero}, M.~I. and {Carrasco}, J.~M. and {Casamiquela}, L. and {Castellani}, M. and {Castro-Ginard}, A. and {Chaoul}, L. and {Charlot}, P. and {Chemin}, L. and {Chiaramida}, V. and {Chiavassa}, A. and {Chornay}, N. and {Comoretto}, G. and {Contursi}, G. and {Cooper}, W.~J. and {Cornez}, T. and {Cowell}, S. and {Crifo}, F. and {Cropper}, M. and {Crosta}, M. and {Crowley}, C. and {Dafonte}, C. and {Dapergolas}, A. and {David}, M. and {David}, P. and {de Laverny}, P. and {De Luise}, F. and {De March}, R.},
        title = "{Gaia Data Release 3. Summary of the content and survey properties}",
      journal = {\aap},
         year = 2023,
        month = jun,
       volume = {674},
          eid = {A1},
        pages = {A1},
          doi = {10.1051/0004-6361/202243940},
archivePrefix = {arXiv},
       eprint = {2208.00211},
 primaryClass = {astro-ph.GA},
       adsurl = {https://ui.adsabs.harvard.edu/abs/2023A&A...674A...1G}
}

@ARTICLE{lindegren+2021,
       author = {{Lindegren}, L. and {Klioner}, S.~A. and {Hern{\'a}ndez}, J. and {Bombrun}, A. and {Ramos-Lerate}, M. and {Steidelm{\"u}ller}, H. and {Bastian}, U. and {Biermann}, M. and {de Torres}, A. and {Gerlach}, E. and {Geyer}, R. and {Hilger}, T. and {Hobbs}, D. and {Lammers}, U. and {McMillan}, P.~J. and {Stephenson}, C.~A. and {Casta{\~n}eda}, J. and {Davidson}, M. and {Fabricius}, C. and {Gracia-Abril}, G. and {Portell}, J. and {Rowell}, N. and {Teyssier}, D. and {Torra}, F. and {Bartolom{\'e}}, S. and {Clotet}, M. and {Garralda}, N. and {Gonz{\'a}lez-Vidal}, J.~J. and {Torra}, J. and {Abbas}, U. and {Altmann}, M. and {Anglada Varela}, E. and {Balaguer-N{\'u}{\~n}ez}, L. and {Balog}, Z. and {Barache}, C. and {Becciani}, U. and {Bernet}, M. and {Bertone}, S. and {Bianchi}, L. and {Bouquillon}, S. and {Brown}, A.~G.~A. and {Bucciarelli}, B. and {Busonero}, D. and {Butkevich}, A.~G. and {Buzzi}, R. and {Cancelliere}, R. and {Carlucci}, T. and {Charlot}, P. and {Cioni}, M.-R.~L. and {Crosta}, M. and {Crowley}, C. and {del Peloso}, E.~F. and {del Pozo}, E. and {Drimmel}, R. and {Esquej}, P. and {Fienga}, A. and {Fraile}, E. and {Gai}, M. and {Garcia-Reinaldos}, M. and {Guerra}, R. and {Hambly}, N.~C. and {Hauser}, M. and {Jan{\ss}en}, K. and {Jordan}, S. and {Kostrzewa-Rutkowska}, Z. and {Lattanzi}, M.~G. and {Liao}, S. and {Licata}, E. and {Lister}, T.~A. and {L{\"o}ffler}, W. and {Marchant}, J.~M. and {Masip}, A. and {Mignard}, F. and {Mints}, A. and {Molina}, D. and {Mora}, A. and {Morbidelli}, R. and {Murphy}, C.~P. and {Pagani}, C. and {Panuzzo}, P. and {Pe{\~n}alosa Esteller}, X. and {Poggio}, E. and {Re Fiorentin}, P. and {Riva}, A. and {Sagrist{\`a} Sell{\'e}s}, A. and {Sanchez Gimenez}, V. and {Sarasso}, M. and {Sciacca}, E. and {Siddiqui}, H.~I. and {Smart}, R.~L. and {Souami}, D. and {Spagna}, A. and {Steele}, I.~A. and {Taris}, F. and {Utrilla}, E. and {van Reeven}, W. and {Vecchiato}, A.},
        title = "{Gaia Early Data Release 3. The astrometric solution}",
      journal = {\aap},
         year = 2021,
        month = may,
       volume = {649},
          eid = {A2},
        pages = {A2},
          doi = {10.1051/0004-6361/202039709},
archivePrefix = {arXiv},
       eprint = {2012.03380},
 primaryClass = {astro-ph.IM},
       adsurl = {https://ui.adsabs.harvard.edu/abs/2021A&A...649A...2L}
}

@ARTICLE{erben_2005,
       author = {{Erben}, T. and {Schirmer}, M. and {Dietrich}, J.~P. and {Cordes}, O. and {Haberzettl}, L. and {Hetterscheidt}, M. and {Hildebrandt}, H. and {Schmithuesen}, O. and {Schneider}, P. and {Simon}, P. and {Deul}, E. and {Hook}, R.~N. and {Kaiser}, N. and {Radovich}, M. and {Benoist}, C. and {Nonino}, M. and {Olsen}, L.~F. and {Prandoni}, I. and {Wichmann}, R. and {Zaggia}, S. and {Bomans}, D. and {Dettmar}, R.~J. and {Miralles}, J.~M.},
        title = "{GaBoDS: The Garching-Bonn Deep Survey. IV. Methods for the image reduction of multi-chip cameras demonstrated on data from the ESO Wide-Field Imager}",
      journal = {Astronomische Nachrichten},
         year = 2005,
        month = jul,
       volume = {326},
       number = {6},
        pages = {432-464},
          doi = {10.1002/asna.200510396},
archivePrefix = {arXiv},
       eprint = {astro-ph/0501144},
 primaryClass = {astro-ph},
       adsurl = {https://ui.adsabs.harvard.edu/abs/2005AN....326..432E}
}

@ARTICLE{ying+2023,
       author = {{Ying}, Jiaqi Martin and {Chaboyer}, Brian and {Boudreaux}, Emily M. and {Slaughter}, Catherine and {Boylan-Kolchin}, Michael and {Weisz}, Daniel},
        title = "{The Absolute Age of M92}",
      journal = {\aj},
         year = 2023,
        month = jul,
       volume = {166},
       number = {1},
          eid = {18},
        pages = {18},
          doi = {10.3847/1538-3881/acd9b1},
archivePrefix = {arXiv},
       eprint = {2306.02180},
 primaryClass = {astro-ph.SR},
       adsurl = {https://ui.adsabs.harvard.edu/abs/2023AJ....166...18Y}
}

@ARTICLE{valcin+2025,
       author = {{Valcin}, David and {Jimenez}, Raul and {Seljak}, Uro{\v{s}} and {Verde}, Licia},
        title = "{The age of the universe with globular clusters. Part III. Gaia distances and hierarchical modeling}",
      journal = {\jcap},
         year = 2025,
        month = oct,
       volume = {2025},
       number = {10},
          eid = {030},
        pages = {030},
          doi = {10.1088/1475-7516/2025/10/030},
archivePrefix = {arXiv},
       eprint = {2503.19481},
 primaryClass = {astro-ph.CO},
       adsurl = {https://ui.adsabs.harvard.edu/abs/2025JCAP...10..030V}
}

@ARTICLE{moya_2026,
       author = {{Moya}, W. Haro and {Bidin}, C. Moni and {Parisi}, M.~C. and {Geisler}, D. and {Bla{\~n}a}, M. and {Villanova}, S. and {Mauro}, F. and {Chen{\'e}}, A.-N. and {Cohen}, R.~E. and {Ram{\'\i}rez Alegr{\'\i}a}, S. and {Contreras Ramos}, R. and {Zoccali}, M. and {Mu{\~n}oz}, C. and {Dias}, B.},
        title = "{Infrared photometry and calcium triplet spectroscopy of the most metal-poor in situ globular cluster VVV-CL001}",
      journal = {\aap},
         year = 2026,
        month = feb,
       volume = {706},
          eid = {A114},
        pages = {A114},
          doi = {10.1051/0004-6361/202556324},
archivePrefix = {arXiv},
       eprint = {2511.13161},
 primaryClass = {astro-ph.GA},
       adsurl = {https://ui.adsabs.harvard.edu/abs/2026A&A...706A.114M}
}

@ARTICLE{Kruijssen2019,
       author = {{Kruijssen}, J.~M. Diederik and {Pfeffer}, Joel L. and {Reina-Campos}, Marta and {Crain}, Robert A. and {Bastian}, Nate},
        title = "{The formation and assembly history of the Milky Way revealed by its globular cluster population}",
      journal = {\mnras},
         year = 2019,
        month = jul,
       volume = {486},
       number = {3},
        pages = {3180-3202},
          doi = {10.1093/mnras/sty1609},
archivePrefix = {arXiv},
       eprint = {1806.05680},
 primaryClass = {astro-ph.GA},
       adsurl = {https://ui.adsabs.harvard.edu/abs/2019MNRAS.486.3180K}
}

@ARTICLE{Gratton2019,
       author = {{Gratton}, Raffaele and {Bragaglia}, Angela and {Carretta}, Eugenio and {D'Orazi}, Valentina and {Lucatello}, Sara and {Sollima}, Antonio},
        title = "{What is a globular cluster? An observational perspective}",
      journal = {\aapr},
         year = 2019,
        month = nov,
       volume = {27},
       number = {1},
          eid = {8},
        pages = {8},
          doi = {10.1007/s00159-019-0119-3},
archivePrefix = {arXiv},
       eprint = {1911.02835},
 primaryClass = {astro-ph.SR},
       adsurl = {https://ui.adsabs.harvard.edu/abs/2019A&ARv..27....8G}
}

@ARTICLE{Valcin2025,
       author = {{Valcin}, David and {Jimenez}, Raul and {Seljak}, Uro{\v{s}} and {Verde}, Licia},
        title = "{The age of the universe with globular clusters. Part III. Gaia distances and hierarchical modeling}",
      journal = {\jcap},
         year = 2025,
        month = oct,
       volume = {2025},
       number = {10},
          eid = {030},
        pages = {030},
          doi = {10.1088/1475-7516/2025/10/030},
archivePrefix = {arXiv},
       eprint = {2503.19481},
 primaryClass = {astro-ph.CO},
       adsurl = {https://ui.adsabs.harvard.edu/abs/2025JCAP...10..030V}
}

@ARTICLE{Bhardwaj2023,
       author = {{Bhardwaj}, Anupam and {Marconi}, Marcella and {Rejkuba}, Marina and {de Grijs}, Richard and {Singh}, Harinder P. and {Braga}, Vittorio F. and {Kanbur}, Shashi and {Ngeow}, Chow-Choong and {Ripepi}, Vincenzo and {Bono}, Giuseppe and {De Somma}, Giulia and {Dall'Ora}, Massimo},
        title = "{Precise Empirical Determination of Metallicity Dependence of Near-infrared Period-Luminosity Relations for RR Lyrae Variables}",
      journal = {\apjl},
         year = 2023,
        month = feb,
       volume = {944},
       number = {2},
          eid = {L51},
        pages = {L51},
          doi = {10.3847/2041-8213/acba7f},
archivePrefix = {arXiv},
       eprint = {2302.05455},
 primaryClass = {astro-ph.SR},
       adsurl = {https://ui.adsabs.harvard.edu/abs/2023ApJ...944L..51B}
}

@ARTICLE{Harris2010,
       author = {{Harris}, William E.},
        title = "{A New Catalog of Globular Clusters in the Milky Way}",
      journal = {arXiv e-prints},
         year = 2010,
        month = dec,
          eid = {arXiv:1012.3224},
        pages = {arXiv:1012.3224},
          doi = {10.48550/arXiv.1012.3224},
archivePrefix = {arXiv},
       eprint = {1012.3224},
 primaryClass = {astro-ph.GA},
       adsurl = {https://ui.adsabs.harvard.edu/abs/2010arXiv1012.3224H}
}

@ARTICLE{Baumgardt2018,
       author = {{Baumgardt}, H. and {Hilker}, M.},
        title = "{A catalogue of masses, structural parameters, and velocity dispersion profiles of 112 Milky Way globular clusters}",
      journal = {\mnras},
         year = 2018,
        month = aug,
       volume = {478},
       number = {2},
        pages = {1520-1557},
          doi = {10.1093/mnras/sty1057},
archivePrefix = {arXiv},
       eprint = {1804.08359},
 primaryClass = {astro-ph.GA},
       adsurl = {https://ui.adsabs.harvard.edu/abs/2018MNRAS.478.1520B}
}

@ARTICLE{Barbuy2018,
       author = {{Barbuy}, B. and {Muniz}, L. and {Ortolani}, S. and {Ernandes}, H. and {Dias}, B. and {Saviane}, I. and {Kerber}, L. and {Bica}, E. and {P{\'e}rez-Villegas}, A. and {Rossi}, L. and et al.},
        title = "{High-resolution abundance analysis of four red giants in the globular cluster NGC 6558★}",
      journal = {\aap},
         year = 2018,
        month = nov,
       volume = {619},
          eid = {A178},
        pages = {A178},
          doi = {10.1051/0004-6361/201833953},
archivePrefix = {arXiv},
       eprint = {1810.11703},
 primaryClass = {astro-ph.SR},
       adsurl = {https://ui.adsabs.harvard.edu/abs/2018A&A...619A.178B}
}

@ARTICLE{Barbuy2006,
       author = {{Barbuy}, B. and {Bica}, E. and {Ortolani}, S. and {Bonatto}, C.},
        title = "{SOAR BVI photometry of the metal-poor bulge globular cluster NGC 6642}",
      journal = {\aap},
         year = 2006,
        month = apr,
       volume = {449},
       number = {3},
        pages = {1019-1024},
          doi = {10.1051/0004-6361:20054073},
archivePrefix = {arXiv},
       eprint = {astro-ph/0511608},
 primaryClass = {astro-ph},
       adsurl = {https://ui.adsabs.harvard.edu/abs/2006A&A...449.1019B}
}

@ARTICLE{Savino2020,
       author = {{Savino}, A. and {Koch}, A. and {Prudil}, Z. and {Kunder}, A. and {Smolec}, R.},
        title = "{The age of the Milky Way inner stellar spheroid from RR Lyrae population synthesis}",
      journal = {\aap},
         year = 2020,
        month = sep,
       volume = {641},
          eid = {A96},
        pages = {A96},
          doi = {10.1051/0004-6361/202038305},
archivePrefix = {arXiv},
       eprint = {2006.12507},
 primaryClass = {astro-ph.GA},
       adsurl = {https://ui.adsabs.harvard.edu/abs/2020A&A...641A..96S}
}

@ARTICLE{CimattiMoresco2023,
       author = {{Cimatti}, Andrea and {Moresco}, Michele},
        title = "{Revisiting the Oldest Stars as Cosmological Probes: New Constraints on the Hubble Constant}",
      journal = {\apj},
         year = 2023,
        month = aug,
       volume = {953},
       number = {2},
          eid = {149},
        pages = {149},
          doi = {10.3847/1538-4357/ace439},
archivePrefix = {arXiv},
       eprint = {2302.07899},
 primaryClass = {astro-ph.CO},
       adsurl = {https://ui.adsabs.harvard.edu/abs/2023ApJ...953..149C}
}

@ARTICLE{Souza2024,
       author = {{Souza}, S.~O. and {Libralato}, M. and {Nardiello}, D. and {Kerber}, L.~O. and {Ortolani}, S. and {P{\'e}rez-Villegas}, A. and {Oliveira}, R.~A.~P. and {Barbuy}, B. and {Bica}, E. and {Griggio}, M. and {Dias}, B.},
        title = "{Combined Gemini-South and HST photometric analysis of the globular cluster NGC 6558: The age of the metal-poor population of the Galactic bulge}",
      journal = {\aap},
         year = 2024,
        month = oct,
       volume = {690},
          eid = {A37},
        pages = {A37},
          doi = {10.1051/0004-6361/202450795},
archivePrefix = {arXiv},
       eprint = {2407.15918},
 primaryClass = {astro-ph.GA},
       adsurl = {https://ui.adsabs.harvard.edu/abs/2024A&A...690A..37S}
}

@ARTICLE{Minitti2010,
       author = {{Minniti}, D. and {Lucas}, P.~W. and {Emerson}, J.~P. and {Saito}, R.~K. and {Hempel}, M. and {Pietrukowicz}, P. and {Ahumada}, A.~V. and {Alonso}, M.~V. and {Alonso-Garcia}, J. and {Arias}, J.~I. and et al.},
        title = "{VISTA Variables in the Via Lactea (VVV): The public ESO near-IR variability survey of the Milky Way}",
      journal = {\na},
         year = 2010,
        month = jul,
       volume = {15},
       number = {5},
        pages = {433-443},
          doi = {10.1016/j.newast.2009.12.002},
archivePrefix = {arXiv},
       eprint = {0912.1056},
 primaryClass = {astro-ph.GA},
       adsurl = {https://ui.adsabs.harvard.edu/abs/2010NewA...15..433M}
}

@ARTICLE{Minniti2011,
       author = {{Minniti}, D. and {Hempel}, M. and {Toledo}, I. and {Ivanov}, V.~D. and {Alonso-Garc{\'\i}a}, J. and {Saito}, R.~K. and {Catelan}, M. and {Geisler}, D. and {Jord{\'a}n}, A. and {Borissova}, J. and et al.},
        title = "{Discovery of VVV CL001. A low-mass globular cluster next to UKS 1 in the direction of the Galactic bulge}",
      journal = {\aap},
         year = 2011,
        month = mar,
       volume = {527},
          eid = {A81},
        pages = {A81},
          doi = {10.1051/0004-6361/201015795},
archivePrefix = {arXiv},
       eprint = {1012.2450},
 primaryClass = {astro-ph.GA},
       adsurl = {https://ui.adsabs.harvard.edu/abs/2011A&A...527A..81M}
}

@ARTICLE{Gran2019,
       author = {{Gran}, F. and {Zoccali}, M. and {Contreras Ramos}, R. and {Valenti}, E. and {Rojas-Arriagada}, A. and {Carballo-Bello}, J.~A. and {Alonso-Garcia}, J. and {Minniti}, D. and {Rejkuba}, M. and {Surot}, F.},
        title = "{Globular cluster candidates in the Galactic bulge: Gaia and VVV view of the latest discoveries}",
      journal = {\aap},
         year = 2019,
        month = aug,
       volume = {628},
          eid = {A45},
        pages = {A45},
          doi = {10.1051/0004-6361/201834986},
archivePrefix = {arXiv},
       eprint = {1904.10872},
 primaryClass = {astro-ph.GA},
       adsurl = {https://ui.adsabs.harvard.edu/abs/2019A&A...628A..45G}
}

@ARTICLE{FT2021,
       author = {{Fern{\'a}ndez-Trincado}, Jos{\'e} G. and {Minniti}, Dante and {Souza}, Stefano O. and {Beers}, Timothy C. and {Geisler}, Doug and {Moni Bidin}, Christian and {Villanova}, Sandro and {Majewski}, Steven R. and {Barbuy}, Beatriz and {P{\'e}rez-Villegas}, Angeles and et al.},
        title = "{VVV CL001: Likely the Most Metal-poor Surviving Globular Cluster in the Inner Galaxy}",
      journal = {\apjl},
         year = 2021,
        month = feb,
       volume = {908},
       number = {2},
          eid = {L42},
        pages = {L42},
          doi = {10.3847/2041-8213/abdf47},
archivePrefix = {arXiv},
       eprint = {2102.01088},
 primaryClass = {astro-ph.GA},
       adsurl = {https://ui.adsabs.harvard.edu/abs/2021ApJ...908L..42F}
}

@ARTICLE{OC2022,
       author = {{Olivares Carvajal}, J. and {Zoccali}, M. and {Rojas-Arriagada}, A. and {Contreras Ramos}, R. and {Gran}, F. and {Valenti}, E. and {Minniti}, J.~H.},
        title = "{Spectroscopic analysis of VVV CL001 cluster with MUSE}",
      journal = {\mnras},
         year = 2022,
        month = jul,
       volume = {513},
       number = {3},
        pages = {3993-4003},
          doi = {10.1093/mnras/stac934},
archivePrefix = {arXiv},
       eprint = {2204.06628},
 primaryClass = {astro-ph.GA},
       adsurl = {https://ui.adsabs.harvard.edu/abs/2022MNRAS.513.3993O}
}

@ARTICLE{MUSE,
       author = {{Weilbacher}, Peter M. and {Palsa}, Ralf and {Streicher}, Ole and {Bacon}, Roland and {Urrutia}, Tanya and {Wisotzki}, Lutz and {Conseil}, Simon and {Husemann}, Bernd and {Jarno}, Aur{\'e}lien and {Kelz}, Andreas and et al.},
        title = "{The data processing pipeline for the MUSE instrument}",
      journal = {\aap},
         year = 2020,
        month = sep,
       volume = {641},
          eid = {A28},
        pages = {A28},
          doi = {10.1051/0004-6361/202037855},
archivePrefix = {arXiv},
       eprint = {2006.08638},
 primaryClass = {astro-ph.IM},
       adsurl = {https://ui.adsabs.harvard.edu/abs/2020A&A...641A..28W}
}

@ARTICLE{APOGEEDR2,
       author = {{J{\"o}nsson}, Henrik and {Holtzman}, Jon A. and {Allende Prieto}, Carlos and {Cunha}, Katia and {Garc{\'\i}a-Hern{\'a}ndez}, D.~A. and {Hasselquist}, Sten and {Masseron}, Thomas and {Osorio}, Yeisson and {Shetrone}, Matthew and {Smith}, Verne and {Stringfellow}, Guy S. and {Bizyaev}, Dmitry and {Edvardsson}, Bengt and {Majewski}, Steven R. and {M{\'e}sz{\'a}ros}, Szabolcs and {Souto}, Diogo and {Zamora}, Olga and {Beaton}, Rachael L. and {Bovy}, Jo and {Donor}, John and {Pinsonneault}, Marc H. and {Poovelil}, Vijith Jacob and {Sobeck}, Jennifer},
        title = "{APOGEE Data and Spectral Analysis from SDSS Data Release 16: Seven Years of Observations Including First Results from APOGEE-South}",
      journal = {\aj},
         year = 2020,
        month = sep,
       volume = {160},
       number = {3},
          eid = {120},
        pages = {120},
          doi = {10.3847/1538-3881/aba592},
archivePrefix = {arXiv},
       eprint = {2007.05537},
 primaryClass = {astro-ph.GA},
       adsurl = {https://ui.adsabs.harvard.edu/abs/2020AJ....160..120J}
}

@ARTICLE{Geisler2023,
       author = {{Geisler}, D. and {Parisi}, M.~C. and {Dias}, B. and {Villanova}, S. and {Mauro}, F. and {Saviane}, I. and {Cohen}, R.~E. and {Moni Bidin}, C. and {Minniti}, D.},
        title = "{Ca triplet metallicities and velocities for 12 globular clusters toward the galactic bulge}",
      journal = {\aap},
         year = 2023,
        month = jan,
       volume = {669},
          eid = {A115},
        pages = {A115},
          doi = {10.1051/0004-6361/202244959},
archivePrefix = {arXiv},
       eprint = {2210.02193},
 primaryClass = {astro-ph.GA},
       adsurl = {https://ui.adsabs.harvard.edu/abs/2023A&A...669A.115G}
}

@ARTICLE{Geisler2025,
       author = {{Geisler}, Doug and {Mu{\~n}oz}, Cesar and {Villanova}, Sandro and {Cohen}, Roger E. and {Minniti}, Dante and {Monachesi}, Antonela and {Majewski}, Steven R. and {Kunder}, Andrea and {Barbuy}, Beatriz and {Cunha}, Katia and et al.},
        title = "{CAPOS: The bulge Cluster APOgee Survey: VIII. Final ASPCAP results for all clusters}",
      journal = {\aap},
         year = 2025,
        month = nov,
       volume = {703},
          eid = {A267},
        pages = {A267},
          doi = {10.1051/0004-6361/202554923},
archivePrefix = {arXiv},
       eprint = {2508.09970},
 primaryClass = {astro-ph.GA},
       adsurl = {https://ui.adsabs.harvard.edu/abs/2025A&A...703A.267G}
}

@ARTICLE{Zoccali2024,
       author = {{Zoccali}, M. and {Quezada}, C. and {Contreras Ramos}, R. and {Valenti}, E. and {Valenzuela-Navarro}, A. and {Olivares Carvajal}, J. and {Rojas Arriagada}, A. and {Minniti}, J.~H. and {Gran}, F. and {De Leo}, M.},
        title = "{VVV catalog of ab-type RR Lyrae in the inner Galactic bulge}",
      journal = {\aap},
         year = 2024,
        month = sep,
       volume = {689},
          eid = {A240},
        pages = {A240},
          doi = {10.1051/0004-6361/202450126},
archivePrefix = {arXiv},
       eprint = {2407.11226},
 primaryClass = {astro-ph.GA},
       adsurl = {https://ui.adsabs.harvard.edu/abs/2024A&A...689A.240Z}
}

@ARTICLE{Olivares2024,
       author = {{Olivares Carvajal}, J. and {Zoccali}, M. and {De Leo}, M. and {Contreras Ramos}, R. and {Quezada}, C. and {Rojas-Arriagada}, {\'A}. and {Valenti}, E. and {Albarrac{\'\i}n}, R. and {Valenzuela Navarro}, {\'A}.},
        title = "{Comparing bulge RR Lyrae stars with bulge giants: Insight from 3D kinematics}",
      journal = {\aap},
         year = 2024,
        month = jul,
       volume = {687},
          eid = {A312},
        pages = {A312},
          doi = {10.1051/0004-6361/202449503},
archivePrefix = {arXiv},
       eprint = {2405.08990},
 primaryClass = {astro-ph.GA},
       adsurl = {https://ui.adsabs.harvard.edu/abs/2024A&A...687A.312O}
}

@ARTICLE{DeLeo2026,
       author = {{De Leo}, M. and {Zoccali}, M. and {Olivares-Carvajal}, J. and {Acosta-Tripailao}, B. and {Gran}, F. and {Contreras-Ramos}, R.},
        title = "{Globular clusters in ORBIT: Complete dynamical characterisation of the Milky Way globular cluster population through updated orbital reconstruction}",
      journal = {\aap},
         year = 2026,
        month = feb,
       volume = {706},
          eid = {A130},
        pages = {A130},
          doi = {10.1051/0004-6361/202556235},
archivePrefix = {arXiv},
       eprint = {2512.06079},
 primaryClass = {astro-ph.GA},
       adsurl = {https://ui.adsabs.harvard.edu/abs/2026A&A...706A.130D}
}

@ARTICLE{Navarro1996,
       author = {{Navarro}, Julio F. and {Frenk}, Carlos S. and {White}, Simon D.~M.},
        title = "{The Structure of Cold Dark Matter Halos}",
      journal = {\apj},
         year = 1996,
        month = may,
       volume = {462},
        pages = {563},
          doi = {10.1086/177173},
archivePrefix = {arXiv},
       eprint = {astro-ph/9508025},
 primaryClass = {astro-ph},
       adsurl = {https://ui.adsabs.harvard.edu/abs/1996ApJ...462..563N}
}

@ARTICLE{Miyamoto1975,
       author = {{Miyamoto}, M. and {Nagai}, R.},
        title = "{Three-Dimensional Models for the Distribution of Mass in Galaxies}",
      journal = {\pasj},
         year = 1975,
        month = dec,
       volume = {27},
       number = {4},
        pages = {533-543},
          doi = {10.1093/pasj/27.4.533},
       adsurl = {https://ui.adsabs.harvard.edu/abs/1975PASJ...27..533M}
}

@ARTICLE{Long1992,
       author = {{Long}, Kevin and {Murali}, Chigurupati},
        title = "{Analytical Potentials for Barred Galaxies}",
      journal = {\apj},
         year = 1992,
        month = sep,
       volume = {397},
        pages = {44},
          doi = {10.1086/171764},
       adsurl = {https://ui.adsabs.harvard.edu/abs/1992ApJ...397...44L}
}

@ARTICLE{Hernquist1990,
       author = {{Hernquist}, Lars},
        title = "{An Analytical Model for Spherical Galaxies and Bulges}",
      journal = {\apj},
         year = 1990,
        month = jun,
       volume = {356},
        pages = {359},
          doi = {10.1086/168845},
       adsurl = {https://ui.adsabs.harvard.edu/abs/1990ApJ...356..359H}
}

@ARTICLE{Astropy2013,
       author = {{Astropy Collaboration} and {Robitaille}, Thomas P. and {Tollerud}, Erik J. and {Greenfield}, Perry and {Droettboom}, Michael and {Bray}, Erik and {Aldcroft}, Tom and {Davis}, Matt and {Ginsburg}, Adam and {Price-Whelan}, Adrian M. and et al.},
        title = "{Astropy: A community Python package for astronomy}",
      journal = {\aap},
         year = 2013,
        month = oct,
       volume = {558},
          eid = {A33},
        pages = {A33},
          doi = {10.1051/0004-6361/201322068},
archivePrefix = {arXiv},
       eprint = {1307.6212},
 primaryClass = {astro-ph.IM},
       adsurl = {https://ui.adsabs.harvard.edu/abs/2013A&A...558A..33A}
}

@ARTICLE{Astropy2018,
       author = {{Astropy Collaboration} and {Price-Whelan}, A.~M. and {Sip{\H{o}}cz}, B.~M. and {G{\"u}nther}, H.~M. and {Lim}, P.~L. and {Crawford}, S.~M. and {Conseil}, S. and {Shupe}, D.~L. and {Craig}, M.~W. and {Dencheva}, N. and et al.},
        title = "{The Astropy Project: Building an Open-science Project and Status of the v2.0 Core Package}",
      journal = {\aj},
         year = 2018,
        month = sep,
       volume = {156},
       number = {3},
          eid = {123},
        pages = {123},
          doi = {10.3847/1538-3881/aabc4f},
archivePrefix = {arXiv},
       eprint = {1801.02634},
 primaryClass = {astro-ph.IM},
       adsurl = {https://ui.adsabs.harvard.edu/abs/2018AJ....156..123A}
}

@ARTICLE{GRAVITY2021,
       author = {{GRAVITY Collaboration} and {Abuter}, R. and {Amorim}, A. and {Baub{\"o}ck}, M. and {Berger}, J.~P. and {Bonnet}, H. and {Brandner}, W. and {Cl{\'e}net}, Y. and {Davies}, R. and {de Zeeuw}, P.~T. and et al.},
        title = "{Improved GRAVITY astrometric accuracy from modeling optical aberrations}",
      journal = {\aap},
         year = 2021,
        month = mar,
       volume = {647},
          eid = {A59},
        pages = {A59},
          doi = {10.1051/0004-6361/202040208},
archivePrefix = {arXiv},
       eprint = {2101.12098},
 primaryClass = {astro-ph.GA},
       adsurl = {https://ui.adsabs.harvard.edu/abs/2021A&A...647A..59G}
}

@ARTICLE{Schonrich2010,
       author = {{Sch{\"o}nrich}, Ralph and {Binney}, James and {Dehnen}, Walter},
        title = "{Local kinematics and the local standard of rest}",
      journal = {\mnras},
         year = 2010,
        month = apr,
       volume = {403},
       number = {4},
        pages = {1829-1833},
          doi = {10.1111/j.1365-2966.2010.16253.x},
archivePrefix = {arXiv},
       eprint = {0912.3693},
 primaryClass = {astro-ph.GA},
       adsurl = {https://ui.adsabs.harvard.edu/abs/2010MNRAS.403.1829S}
}

@ARTICLE{McMillan2017,
       author = {{McMillan}, Paul J.},
        title = "{The mass distribution and gravitational potential of the Milky Way}",
      journal = {\mnras},
         year = 2017,
        month = feb,
       volume = {465},
       number = {1},
        pages = {76-94},
          doi = {10.1093/mnras/stw2759},
archivePrefix = {arXiv},
       eprint = {1608.00971},
 primaryClass = {astro-ph.GA},
       adsurl = {https://ui.adsabs.harvard.edu/abs/2017MNRAS.465...76M}
}

@ARTICLE{Olivares2026,
       author = {{Olivares Carvajal}, J. and {Rojas-Arriagada}, {\'A}. and {Zoccali}, M. and {Acosta-Tripailao}, B. and {De Leo}, M. and {Albarrac{\'\i}n}, R. and {Sanchez-Benavente}, M. and {Valenti}, E. and {Bono}, G. and {Duffau}, S. and et al.},
        title = "{Spectroscopic metallicities and first {\ensuremath{\alpha}}-element abundances of RR Lyrae stars in Baade's Window}",
      journal = {\aap},
         year = 2026,
        month = jul,
       volume = {711},
          eid = {A288},
        pages = {A288},
          doi = {10.1051/0004-6361/202558814},
archivePrefix = {arXiv},
       eprint = {2605.28497},
 primaryClass = {astro-ph.GA},
       adsurl = {https://ui.adsabs.harvard.edu/abs/2026A&A...711A.288O}
}

@ARTICLE{Zoccali2018,
       author = {{Zoccali}, M. and {Valenti}, E. and {Gonzalez}, O.~A.},
        title = "{Weighing the two stellar components of the Galactic bulge}",
      journal = {\aap},
         year = 2018,
        month = oct,
       volume = {618},
          eid = {A147},
        pages = {A147},
          doi = {10.1051/0004-6361/201833147},
archivePrefix = {arXiv},
       eprint = {1807.06377},
 primaryClass = {astro-ph.GA},
       adsurl = {https://ui.adsabs.harvard.edu/abs/2018A&A...618A.147Z}
}

@ARTICLE{Lim2021,
       author = {{Lim}, D. and {Koch-Hansen}, A.~J. and {Chung}, C. and {Johnson}, C.~I. and {Kunder}, A. and {Simion}, I.~T. and {Rich}, R.~M. and {Clarkson}, W.~I. and {Pilachowski}, C.~A. and {Michael}, S. and et al.},
        title = "{Blanco DECam Bulge Survey (BDBS). III. A new view of the double red clump in the Milky Way bulge through luminosity and color distribution}",
      journal = {\aap},
         year = 2021,
        month = mar,
       volume = {647},
          eid = {A34},
        pages = {A34},
          doi = {10.1051/0004-6361/202039955},
archivePrefix = {arXiv},
       eprint = {2101.06269},
 primaryClass = {astro-ph.GA},
       adsurl = {https://ui.adsabs.harvard.edu/abs/2021A&A...647A..34L}
}

@ARTICLE{Hidalgo2018,
       author = {{Hidalgo}, Sebastian L. and {Pietrinferni}, Adriano and {Cassisi}, Santi and {Salaris}, Maurizio and {Mucciarelli}, Alessio and {Savino}, Alessandro and {Aparicio}, Antonio and {Silva Aguirre}, Victor and {Verma}, Kuldeep},
        title = "{The Updated BaSTI Stellar Evolution Models and Isochrones. I. Solar-scaled Calculations}",
      journal = {\apj},
         year = 2018,
        month = apr,
       volume = {856},
       number = {2},
          eid = {125},
        pages = {125},
          doi = {10.3847/1538-4357/aab158},
archivePrefix = {arXiv},
       eprint = {1802.07319},
 primaryClass = {astro-ph.GA},
       adsurl = {https://ui.adsabs.harvard.edu/abs/2018ApJ...856..125H}
}

@ARTICLE{Pietrinferni2021,
       author = {{Pietrinferni}, Adriano and {Hidalgo}, Sebastian and {Cassisi}, Santi and {Salaris}, Maurizio and {Savino}, Alessandro and {Mucciarelli}, Alessio and {Verma}, Kuldeep and {Silva Aguirre}, Victor and {Aparicio}, Antonio and {Ferguson}, Jason W.},
        title = "{Updated BaSTI Stellar Evolution Models and Isochrones. II. {\ensuremath{\alpha}}-enhanced Calculations}",
      journal = {\apj},
         year = 2021,
        month = feb,
       volume = {908},
       number = {1},
          eid = {102},
        pages = {102},
          doi = {10.3847/1538-4357/abd4d5},
archivePrefix = {arXiv},
       eprint = {2012.10085},
 primaryClass = {astro-ph.SR},
       adsurl = {https://ui.adsabs.harvard.edu/abs/2021ApJ...908..102P}
}

@ARTICLE{Tang2014,
       author = {{Tang}, Jing and {Bressan}, Alessandro and {Rosenfield}, Philip and {Slemer}, Alessandra and {Marigo}, Paola and {Girardi}, L{\'e}o and {Bianchi}, Luciana},
        title = "{New PARSEC evolutionary tracks of massive stars at low metallicity: testing canonical stellar evolution in nearby star-forming dwarf galaxies}",
      journal = {\mnras},
         year = 2014,
        month = dec,
       volume = {445},
       number = {4},
        pages = {4287-4305},
          doi = {10.1093/mnras/stu2029},
archivePrefix = {arXiv},
       eprint = {1410.1745},
 primaryClass = {astro-ph.SR},
       adsurl = {https://ui.adsabs.harvard.edu/abs/2014MNRAS.445.4287T}
}

@ARTICLE{Bressan2012,
       author = {{Bressan}, Alessandro and {Marigo}, Paola and {Girardi}, L{\'e}o. and {Salasnich}, Bernardo and {Dal Cero}, Claudia and {Rubele}, Stefano and {Nanni}, Ambra},
        title = "{PARSEC: stellar tracks and isochrones with the PAdova and TRieste Stellar Evolution Code}",
      journal = {\mnras},
         year = 2012,
        month = nov,
       volume = {427},
       number = {1},
        pages = {127-145},
          doi = {10.1111/j.1365-2966.2012.21948.x},
archivePrefix = {arXiv},
       eprint = {1208.4498},
 primaryClass = {astro-ph.SR},
       adsurl = {https://ui.adsabs.harvard.edu/abs/2012MNRAS.427..127B}
}

@ARTICLE{Massari2023,
       author = {{Massari}, Davide and {Aguado-Agelet}, Fernando and {Monelli}, Matteo and {Cassisi}, Santi and {Pancino}, Elena and {Saracino}, Sara and {Gallart}, Carme and {Ruiz-Lara}, Tom{\'a}s and {Fern{\'a}ndez-Alvar}, Emma and {Surot}, Francisco and et al.},
        title = "{Cluster Ages to Reconstruct the Milky Way Assembly (CARMA). I. The final word on the origin of NGC 6388 and NGC 6441}",
      journal = {\aap},
         year = 2023,
        month = dec,
       volume = {680},
          eid = {A20},
        pages = {A20},
          doi = {10.1051/0004-6361/202347289},
archivePrefix = {arXiv},
       eprint = {2310.01495},
 primaryClass = {astro-ph.GA},
       adsurl = {https://ui.adsabs.harvard.edu/abs/2023A&A...680A..20M}
}

@ARTICLE{Cohen2015,
       author = {{Cohen}, Roger E. and {Hempel}, Maren and {Mauro}, Francesco and {Geisler}, Douglas and {Alonso-Garcia}, Javier and {Kinemuchi}, Karen},
        title = "{Wide Field Near-infrared Photometry of 12 Galactic Globular Clusters: Observations Versus Models on the Red Giant Branch}",
      journal = {\aj},
         year = 2015,
        month = dec,
       volume = {150},
       number = {6},
          eid = {176},
        pages = {176},
          doi = {10.1088/0004-6256/150/6/176},
archivePrefix = {arXiv},
       eprint = {1509.01470},
 primaryClass = {astro-ph.SR},
       adsurl = {https://ui.adsabs.harvard.edu/abs/2015AJ....150..176C}
}

@article{Ordenes2026,
	author = {{Ordenes-Huanca, C.} and {Bayo, A.} and {Guarcello, M.} and {Zoccali, M.} and {Rojas-Arriagada, A.} and {Almendros-Abad, V.} and {Andersen, M.} and {Bonito, R.} and {Guzmán, V.} and {Moraux, E.} and {Muzic, K.} and {Prisinzano, L.} and {Sciortino, S.} and {Wright, N. J.}},
	title = {EWOCS-VI: Probing the hidden intermediate-mass population of Westerlund 1},
	DOI= "10.1051/0004-6361/202660600",
	url= "https://doi.org/10.1051/0004-6361/202660600",
	journal = {A\&A},
	year = 2026,
	volume = 712,
	pages = "A193",
}

@MISC{massari+2023,
       author = {{Massari}, Davide and {Bellazzini}, Michele and {Aguado}, Fernando and {Bellini}, Andrea and {Cassisi}, Santi and {Ceccarelli}, Edoardo and {Dalessandro}, Emanuele and {Gallart}, Carme and {Libralato}, Mattia and {Monelli}, Matteo and {Mucciarelli}, Alessio and {Pancino}, Elena and {Salaris}, Maurizio and {Saracino}, Sara},
        title = "{The Hubble Missing Globular Clusters Survey}",
 howpublished = {HST Proposal. Cycle 31, ID. \#17435},
         year = 2023,
        month = aug,
        pages = {17435},
       adsurl = {https://ui.adsabs.harvard.edu/abs/2023hst..prop17435M}
}

@ARTICLE{libralato+2026,
       author = {{Libralato}, M. and {Bellini}, A. and {Massari}, D. and {Bellazzini}, M. and {Aguado-Agelet}, F. and {Cassisi}, S. and {Ceccarelli}, E. and {Dalessandro}, E. and {Dodd}, E. and {Ferraro}, F.~R. and {Gallart}, C. and {Lanzoni}, B. and {Monelli}, M. and {Mucciarelli}, A. and {Pancino}, E. and {Pascale}, R. and {Rosignoli}, L. and {Salaris}, M. and {Saracino}, S. and {Zerbinati}, C.},
        title = "{The Hubble Missing Globular Cluster Survey: III. Astro-photometric catalogs, artificial-star tests, and improved absolute proper motions}",
      journal = {\aap},
         year = 2026,
        month = may,
       volume = {709},
          eid = {A140},
        pages = {A140},
          doi = {10.1051/0004-6361/202659363},
archivePrefix = {arXiv},
       eprint = {2603.25848},
 primaryClass = {astro-ph.GA},
       adsurl = {https://ui.adsabs.harvard.edu/abs/2026A&A...709A.140L}
}

@ARTICLE{Girardi2008,
       author = {{Girardi}, L{\'e}o and {Dalcanton}, Julianne and {Williams}, Benjamin and {de Jong}, Roelof and {Gallart}, Carme and {Monelli}, Matteo and {Groenewegen}, Martin A.~T. and {Holtzman}, Jon A. and {Olsen}, Knut A.~G. and {Seth}, Anil C. and et al.},
        title = "{Revised Bolometric Corrections and Interstellar Extinction Coefficients for the ACS and WFPC2 Photometric Systems}",
      journal = {\pasp},
         year = 2008,
        month = may,
       volume = {120},
       number = {867},
        pages = {583},
          doi = {10.1086/588526},
archivePrefix = {arXiv},
       eprint = {0804.0498},
 primaryClass = {astro-ph},
       adsurl = {https://ui.adsabs.harvard.edu/abs/2008PASP..120..583G}
}

@ARTICLE{Chen2019,
       author = {{Chen}, Yang and {Girardi}, L{\'e}o and {Fu}, Xiaoting and {Bressan}, Alessandro and {Aringer}, Bernhard and {Dal Tio}, Piero and {Pastorelli}, Giada and {Marigo}, Paola and {Costa}, Guglielmo and {Zhang}, Xing},
        title = "{YBC: a stellar bolometric corrections database with variable extinction coefficients. Application to PARSEC isochrones}",
      journal = {\aap},
         year = 2019,
        month = dec,
       volume = {632},
          eid = {A105},
        pages = {A105},
          doi = {10.1051/0004-6361/201936612},
archivePrefix = {arXiv},
       eprint = {1910.09037},
 primaryClass = {astro-ph.SR},
       adsurl = {https://ui.adsabs.harvard.edu/abs/2019A&A...632A.105C}
}

@ARTICLE{Dolphin2000,
       author = {{Dolphin}, Andrew E.},
        title = "{WFPC2 Stellar Photometry with HSTPHOT}",
      journal = {\pasp},
         year = 2000,
        month = oct,
       volume = {112},
       number = {776},
        pages = {1383-1396},
          doi = {10.1086/316630},
archivePrefix = {arXiv},
       eprint = {astro-ph/0006217},
 primaryClass = {astro-ph},
       adsurl = {https://ui.adsabs.harvard.edu/abs/2000PASP..112.1383D}
}

@software{Dolphin2016,
       author = {{Dolphin}, Andrew},
        title = "{DOLPHOT: Stellar photometry}",
 howpublished = {Astrophysics Source Code Library, record ascl:1608.013},
         year = 2016,
        month = aug,
          eid = {ascl:1608.013},
archivePrefix = {ascl},
       eprint = {1608.013},
       adsurl = {https://ui.adsabs.harvard.edu/abs/2016ascl.soft08013D}
}

@inproceedings{taylor2005,
  author = {{Taylor}, M.~B.},
  title = {{TOPCAT \& STIL: Starlink Table/VOTable Processing Software}},
  booktitle = {Astronomical Data Analysis Software and Systems XIV},
  year = {2005},
  editor = {{Shopbell}, P. and {Britton}, M. and {Ebert}, R.},
  series = {Astronomical Society of the Pacific Conference Series},
  volume = {347},
  month = dec,
  pages = {29},
  adsurl = {https://ui.adsabs.harvard.edu/abs/2005ASPC..347...29T},
}

@article{taylor2017,
  author = {{Taylor}, Mark},
  title = {{TOPCAT: Desktop Exploration of Tabular Data for Astronomy and Beyond}},
  journal = {arXiv e-prints},
  year = {2017},
  month = jul,
  eid = {arXiv:1707.02160},
  pages = {arXiv:1707.02160},
  doi = {10.48550/arXiv.1707.02160},
  archiveprefix = {arXiv},
  eprint = {1707.02160},
  primaryclass = {astro-ph.IM},
  adsurl = {https://ui.adsabs.harvard.edu/abs/2017arXiv170702160T},
}

@ARTICLE{Clementini2023,
       author = {{Clementini}, G. and {Ripepi}, V. and {Garofalo}, A. and {Molinaro}, R. and {Muraveva}, T. and {Leccia}, S. and {Rimoldini}, L. and {Holl}, B. and {Jevardat de Fombelle}, G. and {Sartoretti}, P. and et al.},
        title = "{Gaia Data Release 3. Specific processing and validation of all-sky RR Lyrae and Cepheid stars: The RR Lyrae sample}",
      journal = {\aap},
         year = 2023,
        month = jun,
       volume = {674},
          eid = {A18},
        pages = {A18},
          doi = {10.1051/0004-6361/202243964},
archivePrefix = {arXiv},
       eprint = {2206.06278},
 primaryClass = {astro-ph.SR},
       adsurl = {https://ui.adsabs.harvard.edu/abs/2023A&A...674A..18C}
}

@ARTICLE{Soszynski2019,
       author = {{Soszy{\'n}ski}, I. and {Udalski}, A. and {Wrona}, M. and {Szyma{\'n}ski}, M.~K. and {Pietrukowicz}, P. and {Skowron}, J. and {Skowron}, D. and {Poleski}, R. and {Koz{\l}owski}, S. and {Mr{\'o}z}, P. and et al.},
        title = "{Over 78 000 RR Lyrae Stars in the Galactic Bulge and Disk from the OGLE Survey}",
      journal = {\actaa},
         year = 2019,
        month = dec,
       volume = {69},
       number = {4},
        pages = {321-337},
          doi = {10.32023/0001-5237/69.4.2},
archivePrefix = {arXiv},
       eprint = {2001.00025},
 primaryClass = {astro-ph.SR},
       adsurl = {https://ui.adsabs.harvard.edu/abs/2019AcA....69..321S}
}

@ARTICLE{Cadelano2020,
       author = {{Cadelano}, M. and {Saracino}, S. and {Dalessandro}, E. and {Ferraro}, F.~R. and {Lanzoni}, B. and {Massari}, D. and {Pallanca}, C. and {Salaris}, M.},
        title = "{Digging for Relics of the Past: The Ancient and Obscured Bulge Globular Cluster NGC 6256}",
      journal = {\apj},
         year = 2020,
        month = may,
       volume = {895},
       number = {1},
          eid = {54},
        pages = {54},
          doi = {10.3847/1538-4357/ab88b3},
archivePrefix = {arXiv},
       eprint = {2004.06131},
 primaryClass = {astro-ph.GA},
       adsurl = {https://ui.adsabs.harvard.edu/abs/2020ApJ...895...54C}
}

@ARTICLE{Montegriffo2023,
       author = {{Montegriffo}, P. and {De Angeli}, F. and {Andrae}, R. and {Riello}, M. and {Pancino}, E. and {Sanna}, N. and {Bellazzini}, M. and {Evans}, D.~W. and {Carrasco}, J.~M. and {Sordo}, R. and et al.},
        title = "{Gaia Data Release 3. External calibration of BP/RP low-resolution spectroscopic data}",
      journal = {\aap},
         year = 2023,
        month = jun,
       volume = {674},
          eid = {A3},
        pages = {A3},
          doi = {10.1051/0004-6361/202243880},
archivePrefix = {arXiv},
       eprint = {2206.06205},
 primaryClass = {astro-ph.IM},
       adsurl = {https://ui.adsabs.harvard.edu/abs/2023A&A...674A...3M}
}

@ARTICLE{Kerber2019,
       author = {{Kerber}, L.~O. and {Libralato}, M. and {Souza}, S.~O. and {Oliveira}, R.~A.~P. and {Ortolani}, S. and {P{\'e}rez-Villegas}, A. and {Barbuy}, B. and {Dias}, B. and {Bica}, E. and {Nardiello}, D.},
        title = "{A deep view of a fossil relic in the Galactic bulge: the Globular Cluster HP 1}",
      journal = {\mnras},
         year = 2019,
        month = apr,
       volume = {484},
       number = {4},
        pages = {5530-5550},
          doi = {10.1093/mnras/stz003},
archivePrefix = {arXiv},
       eprint = {1901.03721},
 primaryClass = {astro-ph.SR},
       adsurl = {https://ui.adsabs.harvard.edu/abs/2019MNRAS.484.5530K}
}

@ARTICLE{Belokurov2024,
       author = {{Belokurov}, Vasily and {Kravtsov}, Andrey},
        title = "{In-situ versus accreted Milky Way globular clusters: a new classification method and implications for cluster formation}",
      journal = {\mnras},
         year = 2024,
        month = feb,
       volume = {528},
       number = {2},
        pages = {3198-3216},
          doi = {10.1093/mnras/stad3920},
archivePrefix = {arXiv},
       eprint = {2309.15902},
 primaryClass = {astro-ph.GA},
       adsurl = {https://ui.adsabs.harvard.edu/abs/2024MNRAS.528.3198B}
}

@ARTICLE{Ying2025,
       author = {{Ying}, Jiaqi (Martin) and {Chaboyer}, Brian and {Boylan-Kolchin}, Michael and {Weisz}, Daniel R. and {Goebel-Bain}, Rowan},
        title = "{The Absolute Age of Milky Way Globular Clusters}",
      journal = {\apj},
         year = 2025,
        month = jul,
       volume = {987},
       number = {1},
          eid = {52},
        pages = {52},
          doi = {10.3847/1538-4357/add471},
archivePrefix = {arXiv},
       eprint = {2505.02969},
 primaryClass = {astro-ph.GA},
       adsurl = {https://ui.adsabs.harvard.edu/abs/2025ApJ...987...52Y}
}

@ARTICLE{MartinFranch2009,
       author = {{Mar{\'\i}n-Franch}, Antonio and {Aparicio}, Antonio and {Piotto}, Giampaolo and {Rosenberg}, Alfred and {Chaboyer}, Brian and {Sarajedini}, Ata and {Siegel}, Michael and {Anderson}, Jay and {Bedin}, Luigi R. and {Dotter}, Aaron and {Hempel}, Maren and {King}, Ivan and {Majewski}, Steven and {Milone}, Antonino P. and {Paust}, Nathaniel and {Reid}, I. Neill},
        title = "{The ACS Survey of Galactic Globular Clusters. VII. Relative Ages}",
      journal = {\apj},
         year = 2009,
        month = apr,
       volume = {694},
       number = {2},
        pages = {1498-1516},
          doi = {10.1088/0004-637X/694/2/1498},
archivePrefix = {arXiv},
       eprint = {0812.4541},
 primaryClass = {astro-ph},
       adsurl = {https://ui.adsabs.harvard.edu/abs/2009ApJ...694.1498M}
}

@ARTICLE{VandenBerg2013,
       author = {{VandenBerg}, Don A. and {Brogaard}, K. and {Leaman}, R. and {Casagrande}, L.},
        title = "{The Ages of 55 Globular Clusters as Determined Using an Improved \textbackslashDelta V\^HB\_TO Method along with Color-Magnitude Diagram Constraints, and Their Implications for Broader Issues}",
      journal = {\apj},
         year = 2013,
        month = oct,
       volume = {775},
       number = {2},
          eid = {134},
        pages = {134},
          doi = {10.1088/0004-637X/775/2/134},
archivePrefix = {arXiv},
       eprint = {1308.2257},
 primaryClass = {astro-ph.GA},
       adsurl = {https://ui.adsabs.harvard.edu/abs/2013ApJ...775..134V}
}

@ARTICLE{AguadoAgelet2026,
       author = {{Aguado-Agelet}, Fernando and {Zerbinati}, Chiara and {Massari}, Davide and {Monelli}, Matteo and {Cassisi}, Santi and {Fanelli}, Cristiano and {Gallart}, Carme and {Ceccarelli}, Edoardo and {Gonz{\'a}lez Koda}, Yllari Kay and {Ruiz-Lara}, Tom{\'a}s and {Saracino}, Sara and {Salaris}, Maurizio},
        title = "{Cluster Ages to Reconstruct the Milky Way Assembly (CARMA). V. The chronological merger tree of the Milky Way}",
      journal = {arXiv e-prints},
         year = 2026,
        month = aug,
          eid = {arXiv:2608.27611},
        pages = {arXiv:2608.27611},
          doi = {10.48550/arXiv.2608.27611},
archivePrefix = {arXiv},
       eprint = {2608.27611},
 primaryClass = {astro-ph.GA},
       adsurl = {https://ui.adsabs.harvard.edu/abs/2026arXiv260827611A}
}

@ARTICLE{Massari2025,
       author = {{Massari}, D. and {Bellazzini}, M. and {Libralato}, M. and {Bellini}, A. and {Dalessandro}, E. and {Ceccarelli}, E. and {Aguado-Agelet}, F. and {Cassisi}, S. and {Gallart}, C. and {Monelli}, M. and {Mucciarelli}, A. and {Pancino}, E. and {Salaris}, M. and {Saracino}, S. and {Dodd}, E. and {Ferraro}, F.~R. and {Garro}, E.~R. and {Lanzoni}, B. and {Pascale}, R. and {Rosignoli}, L.},
        title = "{The Hubble Missing Globular Cluster Survey: I. Survey overview and the first precise age estimate for ESO452-11 and 2MASS-GC01}",
      journal = {\aap},
         year = 2025,
        month = jun,
       volume = {698},
          eid = {A197},
        pages = {A197},
          doi = {10.1051/0004-6361/202554007},
archivePrefix = {arXiv},
       eprint = {2502.01741},
 primaryClass = {astro-ph.GA},
       adsurl = {https://ui.adsabs.harvard.edu/abs/2025A&A...698A.197M}
}

@ARTICLE{Massari2019,
       author = {{Massari}, D. and {Koppelman}, H.~H. and {Helmi}, A.},
        title = "{Origin of the system of globular clusters in the Milky Way}",
      journal = {\aap},
         year = 2019,
        month = oct,
       volume = {630},
          eid = {L4},
        pages = {L4},
          doi = {10.1051/0004-6361/201936135},
archivePrefix = {arXiv},
       eprint = {1906.08271},
 primaryClass = {astro-ph.GA},
       adsurl = {https://ui.adsabs.harvard.edu/abs/2019A&A...630L...4M}
}

@ARTICLE{Massari2026,
       author = {{Massari}, Davide and {Zerbinati}, Chiara and {Fanelli}, Cristiano and {Helmi}, Amina and {Ceccarelli}, Edoardo and {Aguado-Agelet}, Fernando and {Cassisi}, Santi and {Wempe}, Ewoud and {Monelli}, Matteo and {Bellini}, Andrea and {Callingham}, Thomas M. and {Woudenberg}, Hanneke C. and {Cohen}, Roger E. and {Gallart}, Carme and {Pancino}, Elena and {Saracino}, Sara and {Salaris}, Maurizio and {Mucciarelli}, Alessio},
        title = "{Evidence of a massive accretion event 1.8 billion years before the Gaia-Sausage-Enceladus merger}",
      journal = {Nature Astronomy},
         year = 2026,
        month = aug,
          doi = {10.1038/s41550-026-02931-5},
archivePrefix = {arXiv},
       eprint = {2601.18896},
 primaryClass = {astro-ph.GA},
       adsurl = {https://ui.adsabs.harvard.edu/abs/2026NatAs.tmp..172M}
}

@ARTICLE{Rix2022,
       author = {{Rix}, Hans-Walter and {Chandra}, Vedant and {Andrae}, Ren{\'e} and {Price-Whelan}, Adrian M. and {Weinberg}, David H. and {Conroy}, Charlie and {Fouesneau}, Morgan and {Hogg}, David W. and {De Angeli}, Francesca and {Naidu}, Rohan P. and {Xiang}, Maosheng and {Ruz-Mieres}, Daniela},
        title = "{The Poor Old Heart of the Milky Way}",
      journal = {\apj},
         year = 2022,
        month = dec,
       volume = {941},
       number = {1},
          eid = {45},
        pages = {45},
          doi = {10.3847/1538-4357/ac9e01},
archivePrefix = {arXiv},
       eprint = {2209.02722},
 primaryClass = {astro-ph.GA},
       adsurl = {https://ui.adsabs.harvard.edu/abs/2022ApJ...941...45R}
}

@ARTICLE{Belokurov2022,
       author = {{Belokurov}, Vasily and {Kravtsov}, Andrey},
        title = "{From dawn till disc: Milky Way's turbulent youth revealed by the APOGEE+Gaia data}",
      journal = {\mnras},
         year = 2022,
        month = jul,
       volume = {514},
       number = {1},
        pages = {689-714},
          doi = {10.1093/mnras/stac1267},
archivePrefix = {arXiv},
       eprint = {2203.04980},
 primaryClass = {astro-ph.GA},
       adsurl = {https://ui.adsabs.harvard.edu/abs/2022MNRAS.514..689B}
}

@ARTICLE{ArdenArentsen2024,
       author = {{Ardern-Arentsen}, Anke and {Monari}, Giacomo and {Queiroz}, Anna B.~A. and {Starkenburg}, Else and {Martin}, Nicolas F. and {Chiappini}, Cristina and {Aguado}, David S. and {Belokurov}, Vasily and {Carlberg}, Ray and {Monty}, Stephanie and {Myeong}, GyuChul and {Schultheis}, Mathias and {Sestito}, Federico and {Venn}, Kim A. and {Vitali}, Sara and {Yuan}, Zhen and {Zhang}, Hanyuan and {Buder}, Sven and {Lewis}, Geraint F. and {Oliver}, William H. and {Wan}, Zhen and {Zucker}, Daniel B.},
        title = "{The Pristine Inner Galaxy Survey - VIII. Characterizing the orbital properties of the ancient, very metal-poor inner Milky Way}",
      journal = {\mnras},
         year = 2024,
        month = may,
       volume = {530},
       number = {3},
        pages = {3391-3411},
          doi = {10.1093/mnras/stae1049},
archivePrefix = {arXiv},
       eprint = {2312.03847},
 primaryClass = {astro-ph.GA},
       adsurl = {https://ui.adsabs.harvard.edu/abs/2024MNRAS.530.3391A}
}

@ARTICLE{Vanzella2017,
       author = {{Vanzella}, E. and {Calura}, F. and {Meneghetti}, M. and {Mercurio}, A. and {Castellano}, M. and {Caminha}, G.~B. and {Balestra}, I. and {Rosati}, P. and {Tozzi}, P. and {De Barros}, S. and {Grazian}, A. and {D'Ercole}, A. and {Ciotti}, L. and {Caputi}, K. and {Grillo}, C. and {Merlin}, E. and {Pentericci}, L. and {Fontana}, A. and {Cristiani}, S. and {Coe}, D.},
        title = "{Paving the way for the JWST: witnessing globular cluster formation at z > 3}",
      journal = {\mnras},
         year = 2017,
        month = jun,
       volume = {467},
       number = {4},
        pages = {4304-4321},
          doi = {10.1093/mnras/stx351},
archivePrefix = {arXiv},
       eprint = {1612.01526},
 primaryClass = {astro-ph.GA},
       adsurl = {https://ui.adsabs.harvard.edu/abs/2017MNRAS.467.4304V}
}

@ARTICLE{Adamo2024,
       author = {{Adamo}, Angela and {Bradley}, Larry D. and {Vanzella}, Eros and {Claeyssens}, Ad{\'e}la{\"\i}de and {Welch}, Brian and {Diego}, Jose M. and {Mahler}, Guillaume and {Oguri}, Masamune and {Sharon}, Keren and {Abdurro'uf} and {Hsiao}, Tiger Yu-Yang and {Xu}, Xinfeng and {Messa}, Matteo and {Lassen}, Augusto E. and {Zackrisson}, Erik and {Brammer}, Gabriel and {Coe}, Dan and {Kokorev}, Vasily and {Ricotti}, Massimo and {Zitrin}, Adi and {Fujimoto}, Seiji and {Inoue}, Akio K. and {Resseguier}, Tom and {Rigby}, Jane R. and {Jim{\'e}nez-Teja}, Yolanda and {Windhorst}, Rogier A. and {Hashimoto}, Takuya and {Tamura}, Yoichi},
        title = "{Bound star clusters observed in a lensed galaxy 460 Myr after the Big Bang}",
      journal = {\nat},
         year = 2024,
        month = aug,
       volume = {632},
       number = {8025},
        pages = {513-516},
          doi = {10.1038/s41586-024-07703-7},
archivePrefix = {arXiv},
       eprint = {2401.03224},
 primaryClass = {astro-ph.GA},
       adsurl = {https://ui.adsabs.harvard.edu/abs/2024Natur.632..513A}
}

@ARTICLE{Claeyssens2023,
  author  = {{Claeyssens}, A. and {Adamo}, A. and {Richard}, J. and
             {Mahler}, G. and {Messa}, M. and {Dessauges-Zavadsky}, M.},
  title   = {Star formation at the smallest scales: a {JWST} study of the
             clump populations in {SMACS0723}},
  journal = {Monthly Notices of the Royal Astronomical Society},
  year    = {2023},
  volume  = {520},
  number  = {2},
  pages   = {2180--2203},
  doi     = {10.1093/mnras/stac3791},
  eprint  = {2208.10450},
  archivePrefix = {arXiv}
}

@ARTICLE{Kruijssen2012,
  author  = {{Kruijssen}, J. M. D.},
  title   = {On the fraction of star formation occurring in bound stellar
             clusters},
  journal = {Monthly Notices of the Royal Astronomical Society},
  year    = {2012},
  volume  = {426},
  number  = {4},
  pages   = {3008--3040},
  doi     = {10.1111/j.1365-2966.2012.21923.x},
  eprint  = {1208.2963},
  archivePrefix = {arXiv}
}

@ARTICLE{Kruijssen2015,
  author  = {{Kruijssen}, J. M. D.},
  title   = {Globular clusters as the relics of regular star formation in
             `normal' high-redshift galaxies},
  journal = {Monthly Notices of the Royal Astronomical Society},
  year    = {2015},
  volume  = {454},
  number  = {2},
  pages   = {1658--1686},
  doi     = {10.1093/mnras/stv2026},
  eprint  = {1509.02163},
  archivePrefix = {arXiv}
}

@ARTICLE{Pfeffer2018,
  author  = {{Pfeffer}, J. and {Kruijssen}, J. M. D. and {Crain}, R. A. and
             {Bastian}, N.},
  title   = {The {E-MOSAICS} project: simulating the formation and
             co-evolution of galaxies and their star cluster populations},
  journal = {Monthly Notices of the Royal Astronomical Society},
  year    = {2018},
  volume  = {475},
  number  = {4},
  pages   = {4309--4346},
  doi     = {10.1093/mnras/stx3124}
}

@ARTICLE{Messa2024,
       author = {{Messa}, Matteo and {Dessauges-Zavadsky}, Miroslava and {Adamo}, Angela and {Richard}, Johan and {Claeyssens}, Ad{\'e}la{\"\i}de},
        title = "{Properties of the brightest young stellar clumps in extremely lensed galaxies at redshifts 4 to 5}",
      journal = {\mnras},
         year = 2024,
        month = apr,
       volume = {529},
       number = {3},
        pages = {2162-2179},
          doi = {10.1093/mnras/stae565},
archivePrefix = {arXiv},
       eprint = {2402.14920},
 primaryClass = {astro-ph.GA},
       adsurl = {https://ui.adsabs.harvard.edu/abs/2024MNRAS.529.2162M}
}

@ARTICLE{ReinaCampos2019,
       author = {{Reina-Campos}, Marta and {Kruijssen}, J.~M. Diederik and {Pfeffer}, Joel L. and {Bastian}, Nate and {Crain}, Robert A.},
        title = "{Formation histories of stars, clusters, and globular clusters in the E-MOSAICS simulations}",
      journal = {\mnras},
         year = 2019,
        month = jul,
       volume = {486},
       number = {4},
        pages = {5838-5852},
          doi = {10.1093/mnras/stz1236},
archivePrefix = {arXiv},
       eprint = {1905.02217},
 primaryClass = {astro-ph.GA},
       adsurl = {https://ui.adsabs.harvard.edu/abs/2019MNRAS.486.5838R}
}

@ARTICLE{Li2019,
       author = {{Li}, Hui and {Gnedin}, Oleg Y.},
        title = "{Star cluster formation in cosmological simulations - III. Dynamical and chemical evolution}",
      journal = {\mnras},
         year = 2019,
        month = jul,
       volume = {486},
       number = {3},
        pages = {4030-4043},
          doi = {10.1093/mnras/stz1114},
archivePrefix = {arXiv},
       eprint = {1810.11036},
 primaryClass = {astro-ph.GA},
       adsurl = {https://ui.adsabs.harvard.edu/abs/2019MNRAS.486.4030L}
}

@ARTICLE{Fall2001,
       author = {{Fall}, S. Michael and {Zhang}, Qing},
        title = "{Dynamical Evolution of the Mass Function of Globular Star Clusters}",
      journal = {\apj},
         year = 2001,
        month = nov,
       volume = {561},
       number = {2},
        pages = {751-765},
          doi = {10.1086/323358},
archivePrefix = {arXiv},
       eprint = {astro-ph/0107298},
 primaryClass = {astro-ph},
       adsurl = {https://ui.adsabs.harvard.edu/abs/2001ApJ...561..751F}
}

\begin{appendix}
\onecolumn
\section{Isochrone corrections for the HST wide filters}
\label{app:CMDisochronescorr}
\begin{figure}[ht!]
\sidecaption
\includegraphics[width=10cm]{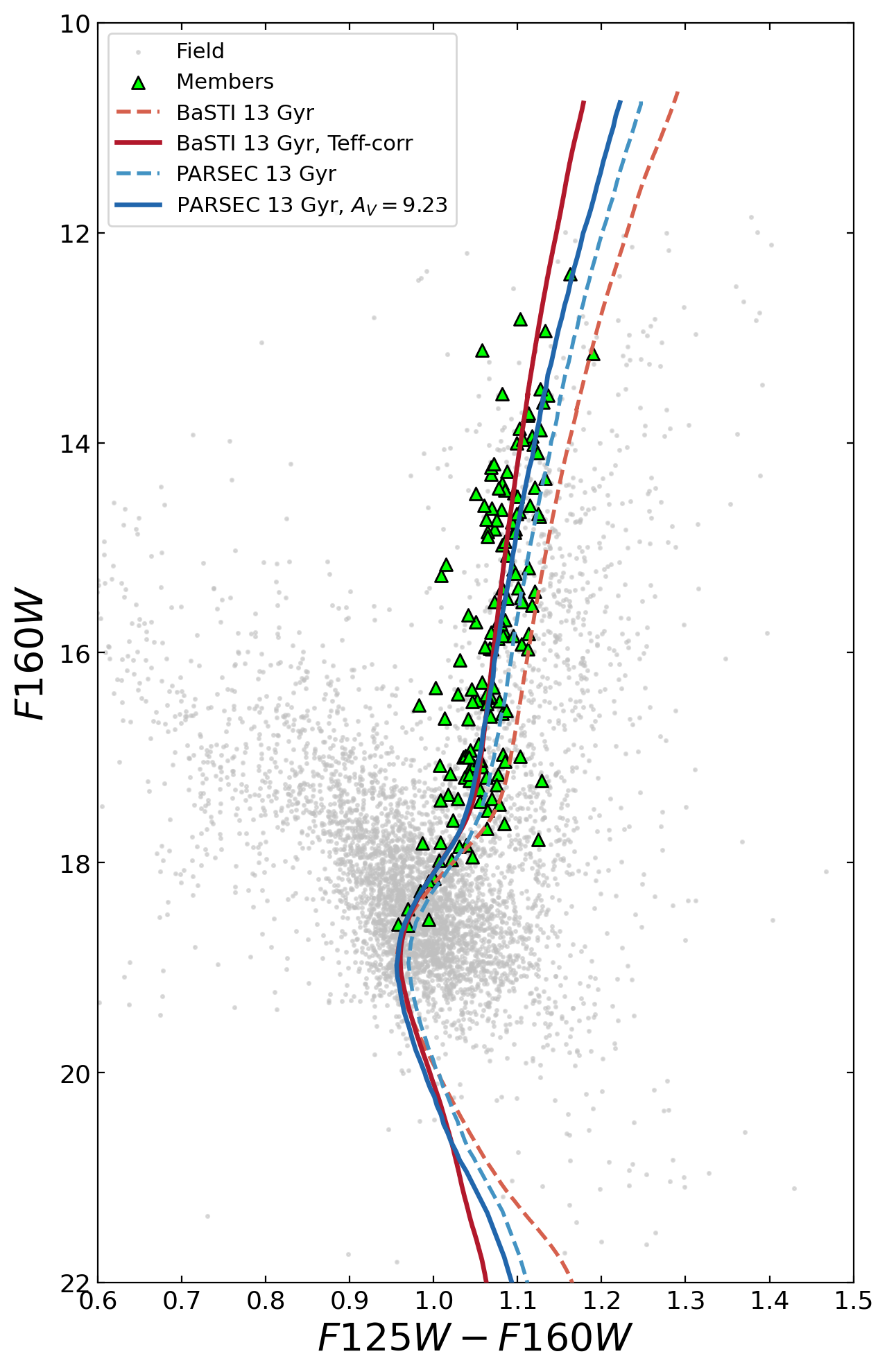}
  \caption{Isochrone corrections for the HST wide filters. The red dashed line is a BaSTI isochrone of 13 Gyr with $\rm [Fe/H]=-2.2$ dex without the $\rm T_{eff}$ correction. The red solid line shows the same isochrone after the correction. The blue dashed line present a PARSEC isochrone of 13 Gyr with $\rm [Fe/H]=-2.2$ dex. The blue solid line show the same PARSEC isochrone with the extinction correction $A_V=9.23$ mag.}
     \label{fig:isocorr}
\end{figure}

To estimate the age of VVV CL001, we used the isochrone-fitting technique, focusing on the TO region. We used the BaSTI~\citep{Hidalgo2018,Pietrinferni2021} and PARSEC~\citep{Bressan2012, Tang2014} isochrone sets independently. In both cases, we detected that the normal isochrones exhibit offsets in the RGB region. Figure~\ref{fig:isocorr} shows that for the BaSTI isochrones, the mean offset was 0.4 mag to the red. The PARSEC isochrone was 0.2 mag redder in the RGB region. These differences were not detected in the NIR FourStar bands. In fact, these offsets were only present in the RGB zone. The main reason for this difference is the differential reddening in this very extincted area of the sky. This affects mostly the colder stars, which, in this case, are on the RGB. For instance, in the TO, the corrected and uncorrected isochrones are very similar. We applied a theoretical $\rm T_{eff}$ correction to the BaSTI isochrones (red solid), as explained in Sect. ~\ref{subsec:age}. In addition, we applied an extinction correction with $A_V=9.23$ mag to the PARSEC isochrones (blue solid). We used these corrected isochrones to perform the isochrone fitting and calculate the age.

\section{Isochrone fitting in the FourStar photometry}
\label{app:CMDFourstar}
\begin{figure}[ht!]
\sidecaption
\includegraphics[width=12cm]{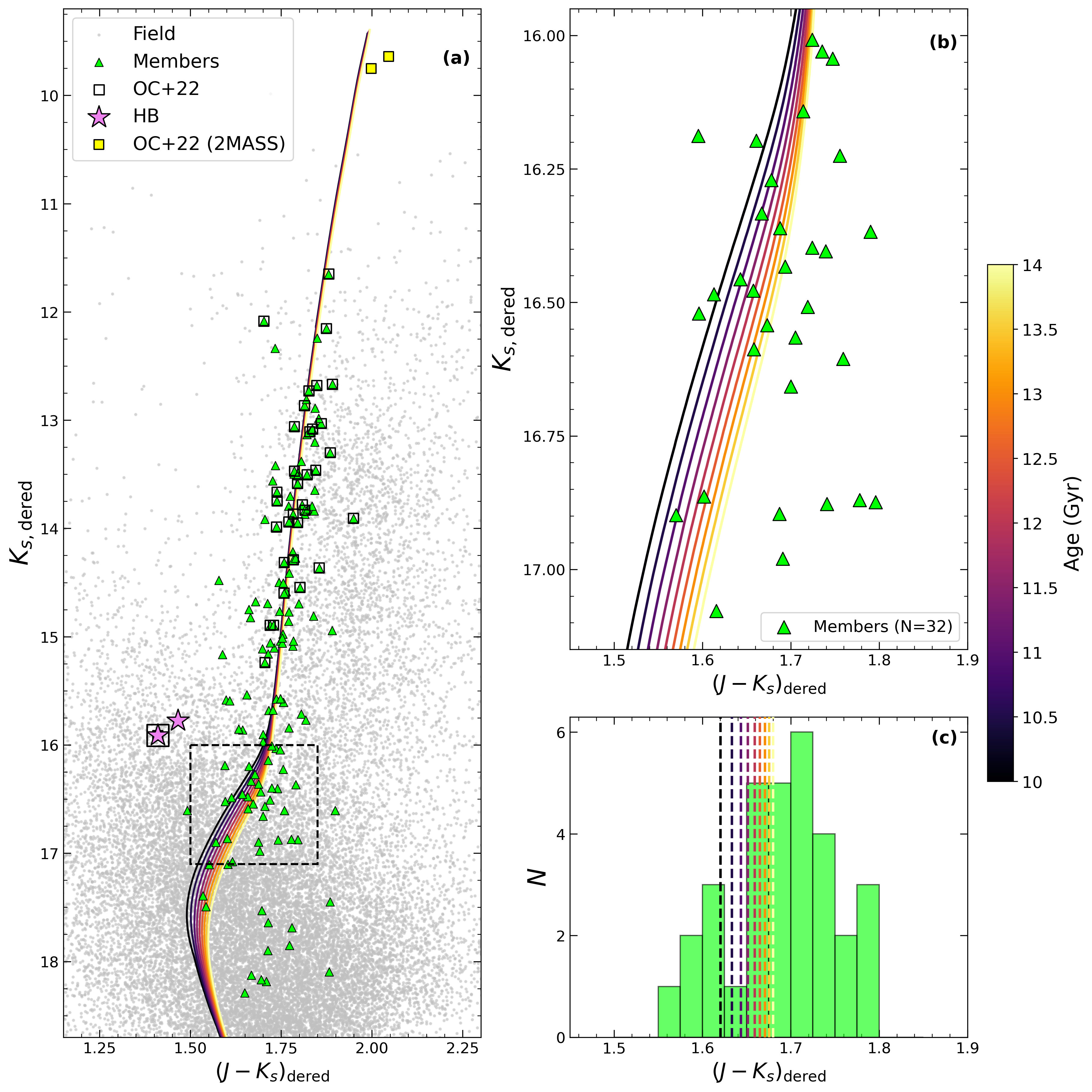}
  \caption{Isochrone fitting around the TO region to calculate the cluster age. \textit{(a):} The complete CMD using the FourStar \J \ and \Ks \ bands. Green triangles are all the selected cluster members. The two violet stars are the HB members. Black empty boxes are the OC22 confirmed members and the two yellow squares are two cluster members with 2MASS photometry from OC22. The gray points are field stars. The BaSTI isochrones are color coded by age. \textit{(b):} Same CMD with a zoom around the TO region. \textit{(c):} The histogram showing the distribution of the TO cluster members. The dashed lines show the isochrone positions in the $F125W-F160W$ color.}
     \label{fig:CMD_JKs}
\end{figure}

Figure ~\ref{fig:CMD_JKs} shows the isochrone fitting using the FourStar \J \ and \Ks \ bands for the cluster members and using the BaSTI isochrones from 10 to 14 Gyr. It is clear that these bands have considerably more error in the deep sources. Moreover, these filters are not the best for separating the isochrones, since they cover similar regions, unlike the HST filters. The cluster's age is limited by the grid to 14 Gyr, so the age derived from these filters is inconclusive. We present the result primarily to demonstrate its consistency with the previous results of this work. Furthermore, it is possible to add the brighter stars of the cluster (yellow squares) with 2MASS photometry from OC22. These stars are very close to the tip of the RGB, indicating that the cluster cannot be much farther away.

\section{Spectra of a HB cluster member}
\label{app:HBmember}
\begin{figure}[ht!]
\sidecaption
\includegraphics[width=12cm]{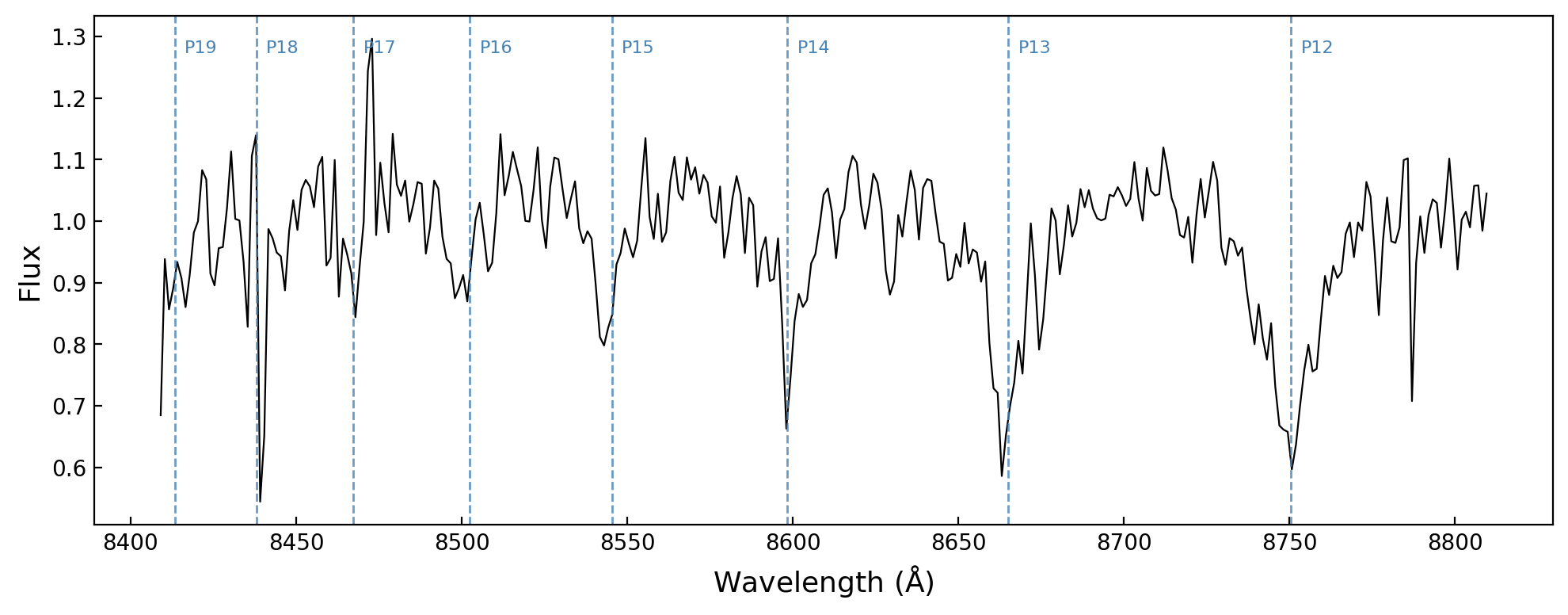}
  \caption{Spectrum of the confirmed HB cluster member. The spectrum came from OC22 and is based on a MUSE datacube with a resolution of $R\sim3000$. The position of the Calcium Triplet is marked with blue vertical lines.}
     \label{fig:spectrum}
\end{figure}

In this work, we confirmed two HB stars as cluster members. Both were confirmed using HDBSCAN and the MCMC method described in Sect.~\ref{sec:members} and classified as high confidence. Figure~\ref{fig:spectrum} presents the spectrum obtained for the star that is also present in OC22. The velocity calculated is approximately -325 km/s, consistent with the cluster mean velocity. Furthermore, the spectrum shows features characteristic of a hot star, as expected for HB stars. The presence of the Paschen lines confirms this result, and they are shown in the figure as blue dashed vertical lines. This test is a valuable double-check for the HB cluster members, since this study is the first to report HB stars in this faint, low-mass GC.

\end{appendix}

\end{document}